\documentclass[preprint,12pt]{elsarticle}

\usepackage{amssymb}
\usepackage{lineno}  
\usepackage{hyperref}
\hypersetup{
  colorlinks=true, 
  linkcolor=black, 
  citecolor=black, 
  urlcolor=blue,   
}
\usepackage{float}
\biboptions{sort&compress}
	
\usepackage{color, colortbl}
	
\definecolor{Gray}{gray}{0.9}

\usepackage{xcolor,makecell,graphicx,amsmath,amssymb,float}
\usepackage{subcaption}
\usepackage[figurename=Fig.]{caption}
\usepackage{svg}
\usepackage{caption}
\newcommand{\orcid}[1]{\href{https://orcid.org/#1}{\includesvg[width=10pt]{orcid}}} 
\makeatletter
\newcommand*{\rom}[1]{\expandafter\@slowromancap\romannumeral #1@}
\makeatother

 \RequirePackage{geometry}
\journal{Journal of Sound and Vibration}

\begin{document}

\begin{frontmatter}



\title{Overcoming stretching and shortening assumptions in Euler-Bernoulli theory using nonlinear Hencky’s beam models: applicable to partly-shortened and partly-stretched beams}



\author{\texorpdfstring{Mohammad Parsa Rezaei $^{a}$}{Mohammad Parsa Rezaei, a}}

\author{\texorpdfstring{Grzegorz Kudra $^{a}$\corref{cor1}\fnref{label2}}{Grzegorz Kudra, a, corresponding author, label2}}
 \ead{grzegorz.kudra@p.lodz.pl}

\author{\texorpdfstring{Mojtaba Ghodsi $^{b}$}{Mojtaba Ghodsi, b}}

\author{\texorpdfstring{Jan Awrejcewicz $^{a}$}{Jan Awrejcewicz, a}}

\affiliation{organization={Department of Automation, Biomechanics and Mechatronics, Lodz University of Technology},
            addressline={\quad  Stefanowski St, 1/15}, 
            city={Lodz},
            postcode={90-537}, 
            state={Lodz},
            country={Poland}}

\affiliation{organization={School of Electrical and Mechanical Engineering, University of Portsmouth},
            addressline={\quad  Anglesea Road}, 
            postcode={PO1 3DJ}, 
            state={Portsmouth},
            country={United Kingdom}}

\begin{abstract}

This paper addresses the challenges of the Euler-Bernoulli beam theory regarding shortening and stretching assumptions. Certain boundary conditions, such as a cantilever with a horizontal spring attached to its end, result in beams that partly shorten or stretch, depending on the spring stiffness. The traditional Euler-Bernoulli beam model may not accurately capture the geometrical nonlinearity in these cases. To address this, nonlinear Hencky's beam models are proposed to describe such conditions.
The validity of these models is assessed against the nonlinear Euler-Bernoulli model using the Galerkin method, with examples including cantilever and clamped-clamped configurations representing shortened and stretched beams. An analysis of a cantilever with a horizontal spring, where stiffness varies, using the nonlinear Hencky's model, indicates that increasing horizontal stiffness stiffens the system. This analysis reveals a transition from softening to linear behavior to hardening near the second resonance frequency, suggesting a bifurcation point.
Despite the computational demands of nonlinear Hencky's models, this study highlights their effectiveness in overcoming the inherent assumptions of stretching and shortening in Euler-Bernoulli beam theory. These models enable a comprehensive nonlinear analysis of partly shortened or stretched beams.
\end{abstract}

\end{frontmatter}


\section{Introduction}
In the field of structural analysis, beam models are fundamental tools. Euler-Bernoulli beam models \cite{LAD2024118545,Bauchau2009,Öchsner2021,WANG2024118585} and Timoshenko beam models \cite{COPETTI2022116920,HAI2024118155,CHALLAMEL2024118602,ZHANG2020115432} are the most commonly used. Among these, Euler-Bernoulli beam models are widely favored due to their simplicity and effectiveness for small deflections \cite{LI20081210}. They are computationally faster, making them more widespread in practical applications. However, the equations of Euler-Bernoulli beam models are highly dependent on boundary conditions, which can cause challenges in accurately modeling certain beam configurations. 

 In 1995, Nyfeh, a prominent researcher in the field, categorized the governing equations of Euler-Bernoulli beams based on the orders of vertical ($W$) and horizontal ($U$) displacements of the beam ($U=O(W^{1 \text{ or } 2})$) depending on the type of boundaries \cite{doi:https://doi.org/10.1002/9783527617586.ch7}. In 2004, he also categorized boundary conditions into two categories  \cite{doi:https://doi.org/10.1002/9783527617562.ch4}:
\begin{enumerate}
    \item Shortened-beam boundary conditions such as Clamped-Free (C-F): Boundaries that one edge is
free or sliding, and no external longitudinal loads are acted.
    \item Stretched-beam boundary conditions such as clamped-clamped (C-C): Boundaries when its edges are prevented from moving.
\end{enumerate}
 Each assumption results in two distinct 2D equations of motion. These can be combined into a single equation; however, this boundary-dependent simplification might fail to conduct a boundary-parametric analysis.

Moreover, in specific boundary conditions, neither of the mentioned assumptions holds true, or it is uncertain which assumptions to accept. According to his book  \cite{doi:https://doi.org/10.1002/9783527617562.ch4}, page 444, when a deformable body is supported in a way that prevents its ends or edges from moving, its midplane experiences nonlinear stretching during transverse vibrations. Therefore, if a cantilever beam has a horizontal spring attached at its free end, it is unknown which assumption to consider. In this case, if the spring's stiffness is zero, the shortening assumption is accurate. Also, if the spring's stiffness is infinite, the stretching assumption is correct. But what if the stiffness is just slightly more than zero? Which assumption should we use then? Partly-shortened or partly-stretched beam? This ambiguity has led to a lack of investigation  \cite{doi:10.1177/1081286517739669,FERNANDEZSAEZ2016107} into a comprehensive nonlinear analysis of some special boundary conditions according to the authors' knowledge.
Despite these challenges, many researchers still use Euler-Bernoulli models based on these assumptions due to their simplicity. To analyze Euler-Bernoulli beams approaches like Galerkin methods and multiple scales are commonly used.

Asymptotic methods \cite{AndrianovBook2024,Andrianov2024} have also been extensively employed in the study of geometrically nonlinear beams. They provide an efficient way to approximate solutions when exact ones are too complex. These methods expand the solution using a small deformation parameter, capturing the main behavior of the system. However, their accuracy depends on the governing equations of motion used in the analysis, such as those from the Euler-Bernoulli theory. 

On the other hand, fewer studies have explored the potential of Hencky's beam models for beam analysis \cite{dell2017mathematical,Challamel2023}. These models offer a more comprehensive framework for bifurcation analysis of boundary conditions, eliminating the need for assumptions about stretching or shortening. This paper seeks to fill this gap by proposing a Nonlinear version of Hencky's beam models that cover Partly-shortened or partly-stretched beams and offer a more accurate bifurcation analysis \cite{Kudra2023}. This model allows us to investigate how variations in critical boundary parameters influence the system's overall behavior and dynamics.

The organization of the paper is as follows: In Section \ref{sec:EulerBernoulli}, we begin by presenting the governing equations of motion for the nonlinear Euler-Bernoulli beam formulation. The subsections \ref{Shortened-beam boundaries} and \ref{Stretched-beam boundaries} provide the non-dimensional governing equations for these specific boundary conditions. We then employ the Galerkin method to discretize the partial differential equations, as detailed in Subsection \ref{Galerkinmethod}. The subsequent Subsection \ref{subsec:NeitherStretchedNorShortened} discusses the challenges faced by the Euler-Bernoulli beam under nonclassical boundary conditions that do not fall into the categories of shortened or stretched beams. In Section \ref{sec:Hencky}, we introduce the nonlinear Hencky’s beam models and derive the governing equations of motion using Lagrange’s equations of the second kind, as outlined in Subsection \ref{sec:Extracting Equations}. The matrix formulation of these governing equations is presented in Subsection \ref{Matrix Formulation}. In Section \ref{Validation}, we assess the validity of Hencky's beam models by comparing them with the nonlinear Euler-Bernoulli model, using examples such as cantilever and clamped-clamped beam configurations representing shortened and stretched beams. Finally, Section \ref{Sec:UniqueApplication} explores a unique application of the nonlinear Hencky’s beam models, analyzing a non-classical support configuration and presenting the results. The paper concludes with a comprehensive summary of the key findings in Section \ref{Sec:Conclusion}.

  \section{Euler-Bernoulli problem formulation} \label{sec:EulerBernoulli}
In this section, we present formulations of the general problem, focusing on a nonlinear Euler-Bernoulli beam structure \cite{doi:https://doi.org/10.1002/9783527617562.ch4}. For simplification, we present C-C and C-F beams as samples of Shortened-beam and Stretched-beam Boundaries, respectively. The physical models of the systems are illustrated in Fig.~\ref{figure1} and Fig.~\ref{figure2}. 

\subsection{Shortened-beam boundaries} \label{Shortened-beam boundaries}
\begin{figure}[H]
\centering
    \includegraphics[width=\linewidth]{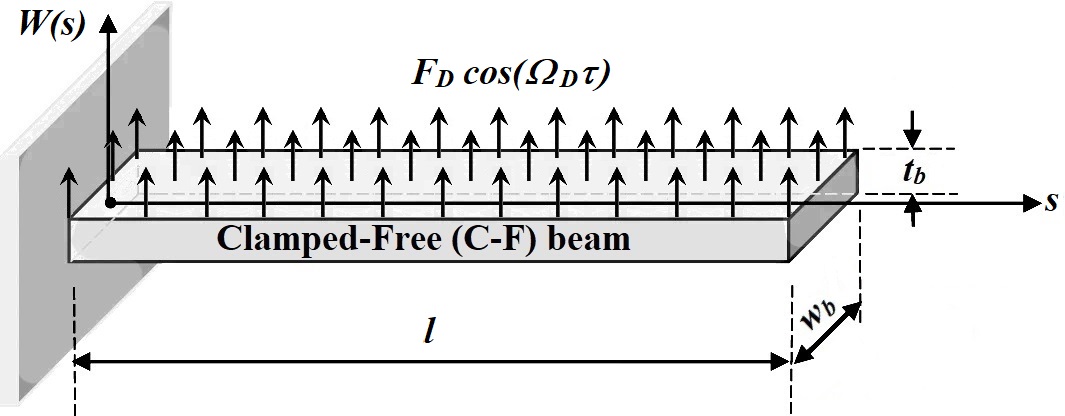}
\caption{Physical model of the clamped-free (C-F) beam system, representing shortened-beam boundaries\label{figure1}}
\end{figure}

When at least one edge is free for longitudinal sliding and no external loads act in the longitudinal direction, assuming that the beam is inextensional holds true. Cantilevers are widely recognized as examples of shortened-beam boundaries. Therefore, we used a cantilever beam as one of the investigated systems as such in Fig.~\ref{figure1}.

In this case, the inextensional assumption confirms that the axial (longitudinal) strain ($e$) of the midplane remains zero. Consequently, the relationship between dimensional horizontal $U$ and transverse $W$ displacement is as follows:
\begin{eqnarray} \label{eq:inextensionality}
\text{For a shortened beam:} \quad e = 0 \quad \Rightarrow \quad \left( 1 + \frac{\partial U}{\partial s} \right)^2 + \left( \frac{\partial W}{\partial s} \right)^2 = 1,
\end{eqnarray}
where \( s \) represents the dimensional horizontal position, and \(\frac{\partial}{\partial s}\) denotes the partial derivative with respect to \( s \).
Eq.~(\ref{eq:inextensionality}) reflects the beam's inextensionality, effectively reducing the number of dependent variables ($U$ and $w$) from two to one ($W$). By applying Newton's second law, expanding the displacement using a third-order Taylor series, and neglecting nonlinear terms related to rotary inertia, the dimensional form of the governing partial differential equation (PDE) is given by \cite{doi:https://doi.org/10.1002/9783527617562.ch4}:

\begin{align} \label{eq:ShorteningWEquationsDimensional}
    \overset{D}{PDE}_{Sh} &= \rho t_b w_b \frac{\partial^2 W}{\partial \tau^2} + EI \frac{\partial^4 W}{\partial s^4} 
    + d_D \frac{\partial}{\partial \tau} \left( \frac{\partial^4 W}{\partial s^4} \right) \nonumber \\
    & + EI \frac{\partial}{\partial s}\left( \frac{\partial W}{\partial s} \left( \frac{\partial^2 W}{\partial s^2} \right)^2 
    + \frac{\partial^3 W}{\partial s^3} \left( \frac{\partial W}{\partial s} \right)^2 \right) \nonumber \\
    & + \frac{1}{2} \frac{\partial}{\partial s} \left[ \frac{\partial W}{\partial s} \int_l^s \rho t_b w_b \hspace{0.1cm} \frac{\partial^2}{\partial \tau^2} \left( \int_0^s 
    \left( \frac{\partial W}{\partial s} \right)^2 ds \right) \right] 
    - F_D \cos(\Omega_D \tau).
\end{align}

Here, $\overset{D}{PDE}_{Sh}$ represents the left-hand side of $\overset{D}{PDE}_{Sh} = 0$. The dimensional time is denoted by \( \tau \), while $\rho$ and $E$ represent the beam's mass density and modulus of elasticity, respectively. As shown in Fig.~\ref{figure1}, the beam's width, length, and thickness are labeled as $w_b$, $l$, and $t_b$. For a rectangular cross-section, the moment of inertia is $I = \frac{1}{12}w_b {t_b}^3$. The term $d_D$ refers to the dimensional internal damping coefficient, while $F_D$ and $\Omega_D$ represent the amplitude and frequency of the vertical base excitation force \cite{clough1993dynamics,GHODSI2019561}.

 To simplify and scale the equations, we introduce the subsequent nondimensional variables.
\begin{equation} \label{eq:nondimensional}
u = \frac{U}{{{l}}},\qquad w = \frac{W}{{{l}}},\qquad x = \frac{s}{l},
\qquad t = \frac{\tau }{T},\qquad T =
\sqrt{\frac{\rho t_b w_b \hspace{0.1cm}l^4}{E I}}.
\end{equation}

 The variables \( u \) and \( w \), here, represent the non-dimensional horizontal and vertical displacements of a point on the beam at non-dimensional horizontal position \( x \) and time \( t \). Hence, based on the non-dimensional variables defined in Eq.~(\ref{eq:nondimensional}), the dimensionless form of the governing equation would be:
\begin{eqnarray} \label{eq:ShorteningWEquations}
 & PDE_{Sh}=\ddot{w} +w^{iv}+d\hspace{0.1cm}\dot{w}^{iv}+(w'{w''}^2+w'''{w'}^2)' & \nonumber \\
&+\frac{1}{2}[ w'\int_1^x \hspace{0.1cm} (\int_0^x {w'}^2 \, dx \ddot{)}  \hspace{0.1cm} ]' -F \cos(\Omega t),&
\end{eqnarray}
where, 
\begin{eqnarray} \label{eq:ShorteningWEquationsParameters}
d= \frac{d_D}{T EI}, \quad F = \frac{F_D l^4}{EI}, \quad \Omega = \Omega_D T.
\end{eqnarray}

$PDE_{Sh}$ defines the left side of $PDE_{Sh} = 0$. The prime (\( '\)) and dot (\( \dot{} \)) notations also indicate differentiation with respect to \(x\) and \(t\), respectively.

\subsection{Stretched-beam boundaries} \label{Stretched-beam boundaries}
\begin{figure}[H]
\centering
    \includegraphics[width=\linewidth]{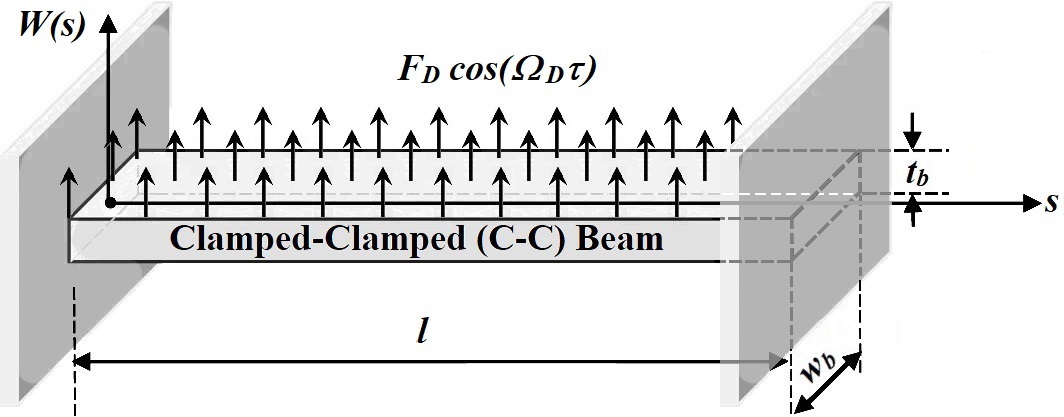}
\caption{Physical model of the clamped-clamped (C-C) beam system, representing stretched-beam boundaries\label{figure2}}
\end{figure}

Nonlinear stretching occurs when the beam is supported in a way that limits the movement of its ends or edges. C-C beams are well-known examples of stretched-beam boundaries. Thus, we included a C-C beam as one of the systems investigated in Fig.~\ref{figure2}.

In this situation, the axial strain ($e$) of the midplane becomes:

\begin{eqnarray} \label{eq:StretchingAxialAtrain}
\text{For stretched beam:}\quad e\neq0 \qquad \text{so}\qquad e=\sqrt{ {(1+\frac{\partial U}{\partial s})}^2+{( \frac{\partial W}{\partial s} )}^2}-1
\end{eqnarray}

Consequently, under the same assumptions as the previous subsection for a clamped-clamped beam, and incorporating Eq.~(\ref{eq:inextensionality}), the dimensional form of the PDE is given by:
\begin{align} \label{eq:StretchingOnedimensionalEquationsDimensional}
    \overset{D}{PDE}_{St} &= \rho t_b w_b \frac{\partial^2 W}{\partial \tau^2} + EI \frac{\partial^4 W}{\partial s^4} 
    + d_D \frac{\partial}{\partial \tau} \left( \frac{\partial^4 W}{\partial s^4} \right) \nonumber \\
    & + EI \left({\frac{\partial^2 W}{\partial s^2}}^3 + 2 \frac{\partial W}{\partial s} \frac{\partial^2 W}{\partial s^2} \frac{\partial^3 W}{\partial s^3} \right) \nonumber \\
    & +\left(\frac{1}{2}\frac{EA}{l}\hspace{0.1cm}\frac{\partial^2 W}{\partial s^2}+\frac{EI}{l}\frac{\partial^4 W}{\partial s^4}\right) \left[\int_0^l \hspace{0.1cm} (\frac{\partial W}{\partial s})^2\, ds \right]- F_D \cos(\Omega_D \tau).
\end{align}

 $\overset{D}{PDE}_{St}$ is the expression that forms the left-hand side of $\overset{D}{PDE}_{St} = 0$. Similarly, using the same non-dimensional variables defined in Eq.~(\ref{eq:nondimensional}), we can obtain the dimensionless form as follows:
\begin{eqnarray} \label{eq:StretchingOnedimensionalEquations}
 &  PDE_{St}=\ddot{w} +w^{iv}+d\hspace{0.1cm}\dot{w}^{iv}+({w''}^3+2 w'w''w''') & \nonumber \\
&+(\frac{1}{2}\frac{12{l}^2}{{t_{b}}^2}\hspace{0.1cm}w''+w^{iv}) [\int_0^1 \hspace{0.1cm} (w')^2\, dx ] -F \cos(\Omega t),&
\end{eqnarray}
where $PDE_{St}$ is the left-hand side of the equation $PDE_{St} = 0$.
\subsection{Solving PDEs via Galerkin method} \label{Galerkinmethod}

Various methods exist for solving partial differential equations (PDEs), such as the Galerkin method \cite{CLARABUT2024117876,Zamanian2015discretization,REZAEI2023104503}  and multiple scales \cite{rezaei2017two}. Among these, we opt for the Galerkin method to discretize PDEs into ordinary differential equations (ODEs), which we subsequently solve using numerical techniques like the Runge-Kutta method.
The fundamental process of the Galerkin method lies in expressing the solution of a PDE in terms of a finite set of basis functions. For PDEs (Eqs.~(\ref{eq:ShorteningWEquations}) and~(\ref{eq:StretchingOnedimensionalEquations})), the solution $w(x, t)$ takes the form:
\begin{eqnarray} \label{eq:GalerkinSolution}
w(x, t) = \sum\limits_{i=1}^{N_m} a_i(t) \psi_i(x),
\end{eqnarray}

Here, $\psi_i(x)$ denotes the $i$-th comparison function, which must satisfy all dynamic and geometric boundary conditions. The total number of modes (comparison functions) is denoted by $N_m$, and $a_i(t)$ represents the unknown generalized coordinates corresponding to these functions. Analogous to the modal expansion method, it is assumed that $\psi_i(x)$ serves as the exact mode shape of the system (the uniform beam). The mathematical expression describing the mode shapes is \cite{DING20122426,Zamanian2015discretization}:
\begin{equation} \label{eq:psi}
\psi_i(x) = A_1 \cosh(\sqrt{\omega_i}x) + A_2 \sinh(\sqrt{\omega_i}x) + A_3 \sin(\sqrt{\omega_i}x) + A_4 \cos(\sqrt{\omega_i}x).
\end{equation}

Here, $A_1$, $A_2$, $A_3$, and $A_4$ are constants, and $\omega_i$ represents the natural frequencies of the uniform beam. When applying boundary conditions to $\psi_i(x)$ and setting the determinant of the coefficient matrix to zero, a set of $\omega_i$ is obtained. Substituting each $\omega_i$ back into the mathematical expression of $\psi_i(x)$ from Eq.~(\ref{eq:psi}), the constants for each mode shape can be determined.
Before applying the Galerkin method, it is beneficial to normalize the mode shapes:
\begin{eqnarray} \label{eq:Normpsi}
 Norm_i =  \rho \: t_b \:w_b \: \sqrt{\int_0^1\psi_i(x)^2 \, dx}, \quad \psi^{Norm}_i(x) = \frac{\psi_i(x)}{Norm_i}.
\end{eqnarray}

Thus, $w(x, t)$ becomes:
\begin{eqnarray} \label{eq:GalerkinSolutionNorm}
w(x, t) = \sum\limits_{i=1}^{N_m} a^{Norm}_i(t)\psi^{Norm}_i(x),
\end{eqnarray}
where $a^{\text{Norm}}_i(t)$ are the unknown generalized coordinates for the normalized mode shapes.
The Galerkin method uses normalized mode shapes as comparison functions. This involves substituting Eq.~(\ref{eq:GalerkinSolutionNorm}) into either Eq.~(\ref{eq:ShorteningWEquations}) or Eq.~(\ref{eq:StretchingOnedimensionalEquations}), multiplying the resulting expression by $\psi^{\text{Norm}}_j(x)$, and integrating over the range $x=0$ to $x=1$:

\begin{enumerate}
    \item \text{For shortened-beam boundaries}: 
    \begin{align}
\label{eq:ODEShortening}
        ODE_j = \int_0^1 \psi^{\text{Norm}}_j(x) \, PDE_{\text{Sh}} \, dx.
    \end{align}

    \item \text{For stretched-beam boundaries}:
    \begin{align}
\label{eq:ODEStretching}
        ODE_j = \int_0^1 \psi^{\text{Norm}}_j(x) \, PDE_{\text{St}} \, dx.
    \end{align}
\end{enumerate}

The term $ODE_j$ denotes the left-hand side of the equation $ODE_j = 0$. The dependence of ordinary differential equations (ODEs) and their solutions on boundary conditions, as indicated by Eqs.~(\ref{eq:ODEShortening}) and~(\ref{eq:ODEStretching}), makes it nearly impossible to carry out boundary-parametric analysis. Finally, these ODEs are tackled numerically, typically utilizing methods like the Runge-Kutta 45 algorithm, implemented in software platforms such as \textsc{Matlab}, to obtain the solution.

\subsection{Partly-shortened or partly-stretched beam boundaries} \label{subsec:NeitherStretchedNorShortened}
\begin{figure}[H]
\centering
\includegraphics[width=  \linewidth]{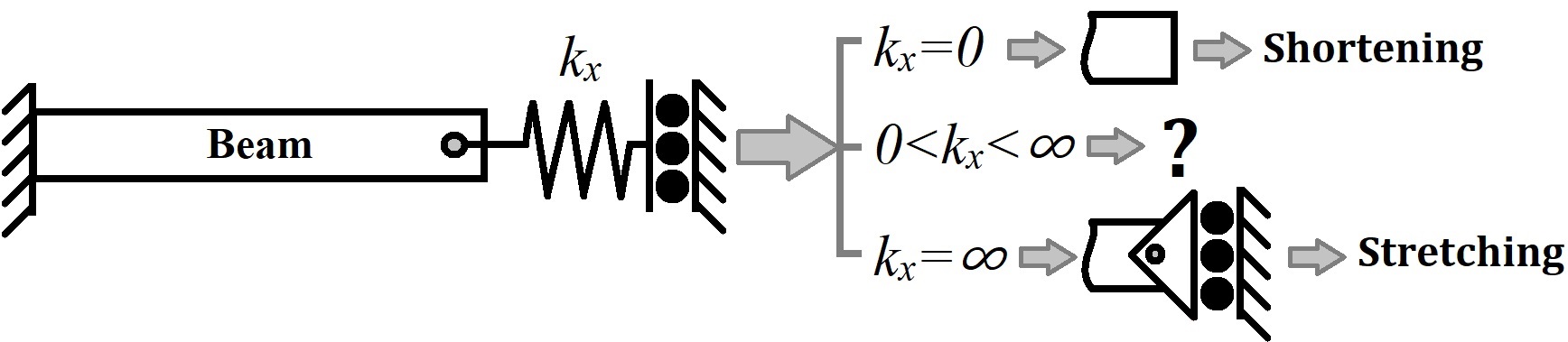}
\caption{Configuration example of Partly-Shortened or Partly-stretched beam boundaries. \label{figure3}}
\end{figure}

In the context of Euler-Bernoulli beam theory, the assumptions of shortening and stretching are fundamental to the model in nonlinear analysis. However, there are boundary conditions where these assumptions may not hold true. One example is a non-classical support configuration shown in Fig.~\ref{figure3}, where the right end combines spring support ($0 \leq k_x \leq \infty$), pinned support, and vertical roller support.
\begin{itemize}
    \item When the spring stiffness is zero ($k_x = 0$), the beam acts like a cantilever (shortening assumption applies, resulting in Eq.~(\ref{eq:ShorteningWEquations})).
    \item When the spring stiffness is infinite ($k_x = \infty$), the beam behaves as a clamped-vertically rolling support (stretching assumption applies).
\end{itemize}
For boundary conditions where $0 < k_x < \infty$, Euler-Bernoulli models fall short because the beams are partly shortened or stretched, depending on the value of $k_x$. They cannot handle the transition from zero to infinite stiffness in a nonlinear context. Nonlinear Hencky beam models overcome this limitation. We will explain these models in Section~\ref{sec:Hencky}, and present results for the example boundary conditions shown in Fig.~\ref{figure3} in Section~\ref{Sec:UniqueApplication}.

\section{Nonlinear Hencky’s beam models}\label{sec:Hencky}
Challamel (2023) compared three types of Hencky beam models: with lumped masses at the joint, lumped masses at the middle of the rigid segment, and distributed mass along the rigid segment, showing that each has its benefits. However, our model differs and is designed to simultaneously cover horizontal, vertical, and angular displacements, improve convergence, model arbitrary boundary conditions at the right end, and account for the full nonlinearity of the beam. The general concepts of the proposed nonlinear Hencky model are as follows (see Fig.~\ref{figure4}):
\begin{enumerate}
\item The dynamical system has the form of a planar horizontal multi-pendulum,  i.e. the thin rods are connected rotationally at their ends, with the first pendulum rotationally suspended at the left end of the system. Therefore the system position is uniquely defined by angular positions ($\phi_i$) of the rods, being also generalized coordinates in Lagrange equations of the second kind used in mathematical modeling of the system.
\item The continuous beam is discretized and the lumped pendulum system is constructed in the following way:
\begin{enumerate}
\item The whole beam is divided into $n$ equal segments;
\item A rotational joint with appropriately chosen (depending on flexural stiffness and internal damping of the beam) rotational spring and damper is placed in the middle of each segment.
\end{enumerate}
The result is a system composed of $n+1$ pendulums, numbered from 0 to $n$. Note that the lengths of the pendulum links are equal, except for the $0^{th}$ and $n$-th pendulums, which are half the length of the others. The $0^{th}$ pendulum on its left end is rotationally joined to the support. Depending on the boundary condition on the left end of the beam, the $0^th$ pendulum can be part of the dynamical system (pinned left end of the beam) or can be fixed at zero position (clamped left end of the beam) and eliminated from the system. In this work, only the second case will be used, so the system is finally composed of $n$ pendulums, numbered from 1 to $n$.
\item General or arbitrary boundary conditions at the right end are modeled by attaching three springs: horizontal (of stiffness $k_x$), vertical ($k_y$), and angular ($k_{\phi}$). 
\item To model the longitudinal flexibility of the beam and cover the full geometric non-linearity of Euler-Bernoulli beams, we consider an additional longitudinal spring ($k_{st}$) connected in series with the horizontal spring modeling boundary conditions ($k_x$) on the right end of the beam.
\end{enumerate}

\subsection{Extracting Equations} \label{sec:Extracting Equations}
 The assumptions made in the previous section~\ref{sec:EulerBernoulli}, such as the negligible influence of gravity, still apply here.
\begin{figure}[H]
\centering
\includegraphics[width= 1 \linewidth]{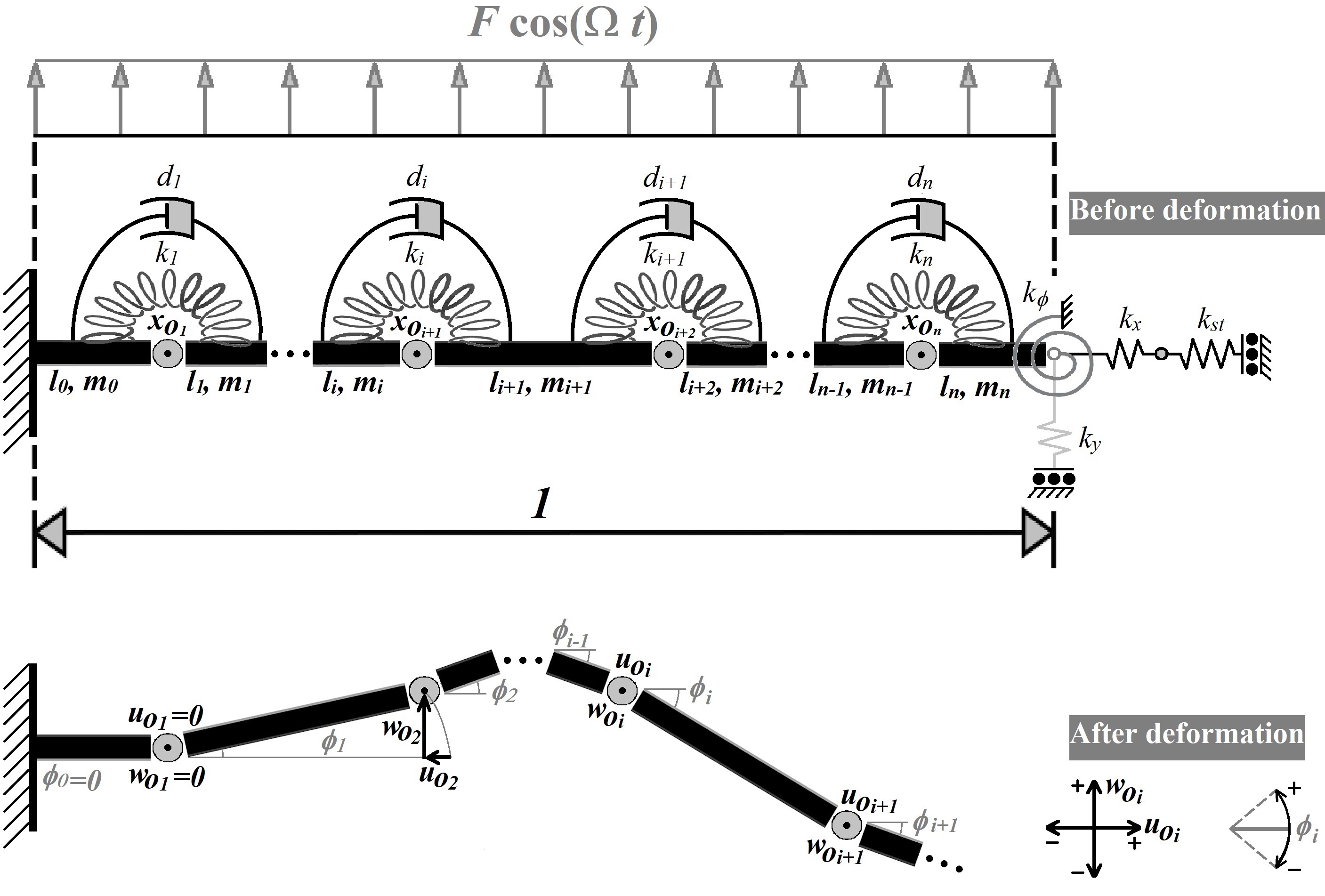}
\caption{Nonlinear Hencky model: system of $n$ pendulums with masses $m_i$ and springs $k_i$. The $1^{\text{st}}$ and $n^{\text{th}}$ pendulums are half-length. Right end springs: longitudinal ($k_{st}$), horizontal ($k_x$), vertical ($k_y$), and angular ($k_\phi$). \label{figure4}}
\end{figure} 

Figure~\ref{figure4} illustrates the beam divided into $n+1$ elements, with pendulums labeled starting from $i=0$. The first pendulum, at $i=0$, is fixed by clamping at the left end.  All parameters depicted are non-dimensional, hence the beam's length is normalized to one. Therefore, the dynamical system ultimately consists of $n$ pendulums. The element lengths are:
\begin{equation} \label{eq:ElementsLength}
l_0=\frac{1}{2} \hspace{0.1cm} l_e, \hspace{0.5cm} l_n=\frac{1}{2}\hspace{0.1cm} l_e,  \hspace{0.5cm}l_i=l_e=\frac{1}{n} \hspace{1cm} i=1...n-1. 
\end{equation}
where  $l_i$ are non-dimensional length of the $i^{th}$ element. It should be highlighted that all parameters of model formulations in this section are non-dimensional. 
Using Lagrange's equations of the second kind:

\begin{equation}\label{eq:lagrange}
\frac{d}{dt}\left(\frac{\partial T}{\partial \dot \phi_i}\right) - \frac{\partial T}{\partial \phi_i} + \frac{\partial V}{\partial \phi_i} + \frac{\partial R}{\partial \dot \phi_i} = Q_i, \hspace{0.2cm} i=1...n.
\end{equation}

The ordinary differential equations are obtained for this structure \cite{kudra2024mathematical}. In these equations, $\phi_i$ represents generalized coordinates that are the angular position of the $i^{th}$ element. $T$, $V$, and $R$ respectively denote kinetic energy, potential energy, and Rayleigh function. There are also generalized forces that act on the system ($Q_i$).

According to Figure~\ref{figure4}, in Henckey's beam model, the horizontal displacement and vertical displacement at the mass center of the $i^{th}$ link, denoted by  ${u_C}_i$ and ${w _C}_i$, are:
\begin{align}
{u_C}_i &=\sum_{j=0}^{i-1} l_j \cos{\phi_j} + e_i \cos{\phi_i}-  \sum_{j=0}^{i-1} l_j - e_i , \quad e_i = \frac{l_i}{2}, \quad \phi_0=0\label{eq:PositionuC}\\
{w _C}_i &= \sum_{j=0}^{i-1} l_j \sin{\phi_j} + e_i \sin{\phi_i}\label{eq:PositionwC}
\end{align}

Let  $x_{O_i}$ be the initial horizontal position at the joint of the $i$-th link (before deformation). It is given by:
\begin{flalign}
\label{eq:PositionxO}
x_{O_i} = \sum_{j=0}^{i} l_j.
\end{flalign}

Then, the horizontal and vertical displacements at the joint of the $i$-th link, denoted by ${u_O}_i$ and ${w_O}_i$, are given by:
\begin{subequations}
    \begin{flalign}
\label{eq:PositionuO}
{u_O}_i &= {u_C}_i + e_i \cos{\phi_i},& \\
{w_O}_i &= {w_C}_i + e_i \sin{\phi_i}. &
\end{flalign}
\end{subequations}

Then, assuming full nonlinearity, the kinetic energy $T$ can be expressed as:
\begin{eqnarray}\label{eq:KineticEnergy}
T &= \frac{1}{2} \sum_{i=1}^{n} \left\{ m_i \left[(\dot{w _C}_i)^2 + (\dot{u_C}_i)^2\right] + B_i (\dot{\phi}_i)^2 \right\},\qquad \phi_0=0, 
\end{eqnarray}
where $m_i$ and $B_i$ represent the non-dimensional mass and mass moment of inertia with respect to the mass center of the $i^{th}$ element,  respectively. These non-dimensional parameters, in connection with the non-dimensional variables in  Eq.~(\ref{eq:nondimensional}), can be derived directly from their dimensional counterparts. By comparing the dimensional parameters in Eqs.~(\ref{eq:ShorteningWEquationsDimensional}) and~(\ref{eq:StretchingOnedimensionalEquationsDimensional}) with the non-dimensional parameters in Eqs.~(\ref{eq:ShorteningWEquations}) and~(\ref{eq:StretchingOnedimensionalEquations}), the corresponding non-dimensional forms of any parameter can be found. For example, from this comparison, one can see that the non-dimensional form of $\rho t_b w_b$ is $1$. Therefore:
\begin{equation}
\label{eq:KineticEnergyParameters1}
m_i =l_i, \quad B_i = \frac{m_i (l_i)^2}{12}.
\end{equation}

The Rayleigh function $R$ in Eq.~\ref{eq:lagrange}, representing dissipative forces, is given by:
\begin{equation}
\label{eq:Rayleigh}
R = \frac{1}{2} \sum_{i=1}^{n} d_i (\dot{\phi}_i - \dot{\phi}_{i-1})^2
\end{equation}

Thus, the beam's internal damping is modeled using $n$ rotational dampers, each characterized by the non-dimensional internal damping coefficient $d_i$. The dimensional internal damping coefficient at the $i^{th}$ joint is obtained by dividing the total dimensional internal damping coefficient of the beam ($d_D$) by the length of the $i^{th}$ element. By comparing Eqs.~(\ref{eq:ShorteningWEquations}),~(\ref{eq:ShorteningWEquationsDimensional}),~(\ref{eq:StretchingOnedimensionalEquations}), and~(\ref{eq:StretchingOnedimensionalEquationsDimensional}), it follows that \( d_i = \frac{d}{l_i} = dn \). 

Lagrange's equations involve calculating the potential energy of the system and its associated parameters using the following equations:
\begin{eqnarray}\label{eq:PotentialEnergy}
&V=\frac{1}{2} [\sum\limits_{i=1}^{n} \hspace{0.1 cm} k_i (\phi_i-\phi_{i-1})^2]+\frac{1}{2}k^{Total}_{x} (u_{end})^2+\frac{1}{2}k_y (w_{end})^2+\frac{1}{2}k_\phi (\phi_{n})^2,    
\end{eqnarray} 
where $u_{end}$ and $w_{end}$ are horizontal and vertical displacements at the end of the beam:
 \begin{subequations}
    \begin{flalign} \label{eq:DeflectionAtEnd}
&u_{end}=\sum\limits_{i=0}^{n} l_i \cos \phi_{i} \: -1,&\\&w_{end}=\sum\limits_{i=0}^{n} l_i \sin \phi_{i}, &
    \end{flalign}
\end{subequations}
and where  $k_{i}$ are the non-dimensional stiffness of rotational springs due to the flexural stiffness along the beam. The stiffness of each rotational spring is obtained by dividing the flexural stiffness by the length of the $i^{th}$ element. A comparison of Eqs.~(\ref{eq:ShorteningWEquations}),~(\ref{eq:ShorteningWEquationsDimensional}),~(\ref{eq:StretchingOnedimensionalEquations}), and~(\ref{eq:StretchingOnedimensionalEquationsDimensional}) reveals that the non-dimensional form of \( EI \) is equal to \( 1 \), indicating that the stiffness for the \( i^{th} \) rotational spring is given by \( k_{i} = \frac{1}{l_i} = n \).

As mentioned, the boundary at the right end is modeled with three springs: one for horizontal movement (${k}_{x}$), one for vertical movement (${k}_{y}$), and one for angular movement (${k}_{\phi}$). Additionally, a longitudinal spring with non-dimensional stiffness \( k_{st} \) accounts for the geometric nonlinearity associated with the beam's axial stiffness, derived from the dimensional form \( \frac{EA}{l} \). By comparing Eqs.~(\ref{eq:StretchingOnedimensionalEquations}) and~(\ref{eq:StretchingOnedimensionalEquationsDimensional}), it is concluded that the non-dimensional form of \( \frac{EA}{l} \) can be expressed as \( \frac{12 l^2}{{t_b}^2} \). Therefore:
\begin{equation}\label{eq:StretchingStiffness}
k_{st}=\frac{12 l^2}{{t_b}^2}.     
\end{equation} 

This spring accounts for the longitudinal flexibility and captures the full geometric nonlinearity of Euler-Bernoulli beams. It is attached in series with the horizontal spring due to the stiffness of the right boundary (see Eq.~(\ref{eq:TotalHorizontalStiffness})). Therefore, the total horizontal stiffness coefficient is ${k}_{x}^{Total}$ which is calculated  as follows:
\begin{equation}\label{eq:TotalHorizontalStiffness}
k^{Total}_{x}=\frac{k_x\hspace{0.2cm}k_{st}}{k_x+k_{st}}.     
\end{equation}  

Eq.~(\ref{eq:TotalHorizontalStiffness}) reveals how Nonlinear Henckey's models overcome the Stretching and shortening Assumptions of the Euler-Bernoulli theory. When dealing with shortened boundaries, such as a cantilever, there is no horizontal movement restriction at the right end. Consequently, the horizontal stiffness (${k}_{x}=0$) at the right end is zero, resulting in a total horizontal stiffness of zero (${k}_{x}^{Total}=0$), thereby eliminating the influence of stretching stiffness (${k}_{st}$). Conversely, when dealing with stretched boundaries, there are restrictions on horizontal movement, leading to a non-zero horizontal stiffness at the right end (${k}_{x}\neq 0$). This allows the influence of stretching stiffness to emerge in the total horizontal stiffness. In other words:
\begin{subequations} \label{eq:OvercomingStretchinShortening}
\begin{flalign}
&\textbf{For fully shortened-beam boundaries (e.g., C-F): }& \nonumber \\
& k_x=0 \Rightarrow \quad k^{Total}_{x}=0 \Rightarrow \quad k_{st} \quad \text{has no influence}&\\
&\textbf{For other boundaries (e.g., C-C): }& \nonumber \\
& k_x \neq 0 \Rightarrow \quad k^{Total}_{x} \neq 0 \Rightarrow \quad k_{st} \quad \text{has influence}&
 \end{flalign}
\end{subequations}

The remaining terms in Eq.~(\ref{eq:lagrange}) represent the generalized external forces ($Q_i$), which account for equivalent vertical base excitation forces acting at the mass centers. These forces are defined as:
\begin{eqnarray}\label{eq:BaseExcitationForce}
 Q_i = F \cos(\Omega t) \sum\limits_{j=1}^{n} m_j \: \frac{\partial {w _C}_{_j}}{\partial \phi_i}.
\end{eqnarray}

They mimic inertia forces on the mass centers, representing vertical base excitation forces acting across the entire beam.

\subsection{Matrix Formulation} \label{Matrix Formulation}
We express the equations (ODEs) of motions for our general system in matrix form for better clarity and computational implementation:
\begin{equation}
\label{eq:HenckeyEquationsMatrixForm}
\mathbf{M} \cdot {\boldsymbol{\ddot \phi}} +  \mathbf{K} \cdot( \boldsymbol{\phi} + d \hspace{0.1cm}  {\boldsymbol{\dot \phi}})+\mathbf{N} \cdot  {\boldsymbol{\dot \phi}}^2+ \mathbf{K}_{B} \cdot \boldsymbol{\phi} = \mathbf{F}
\end{equation}
where $\boldsymbol{\phi}$, $\boldsymbol{\dot \phi}$, and $\boldsymbol{\ddot \phi}$ are vectors representing the generalized coordinates, the first derivatives, and the second derivatives of the generalized coordinates of the system, as given by:
\begin{eqnarray}\label{eq:PhiVector}
 \boldsymbol{\phi} = 
\begin{bmatrix}
\phi_1 & \cdots & \phi_n
\end{bmatrix}^\top,
 \quad  \boldsymbol{\dot \phi} = 
\begin{bmatrix}
\dot \phi_1 & \cdots & \dot \phi_n
\end{bmatrix}^\top,\\
 \quad  {\boldsymbol{\dot \phi}}^2 = 
\begin{bmatrix}
{\dot \phi_1}^2 & \cdots & {\dot \phi_n}^2
\end{bmatrix}^\top,
 \quad  \boldsymbol{\ddot \phi} = 
\begin{bmatrix}
\ddot \phi_1 & \cdots & \ddot \phi_n
\end{bmatrix}^\top,
\end{eqnarray}
and where $\mathbf{M}$, $\mathbf{K}$, $\mathbf{N}$, $\mathbf{K_B}$ and $\mathbf{F}$ are matrices and vectors defined below.
To simplify the matrices for presentation purposes, we introduce indexing parameters:
\begin{subequations}\label{eq:SijAndCij}
\begin{align}
&{s}_{i} = \sin(\phi_i(t)), \qquad {s}_{i,j} = \sin(\phi_i(t) - \phi_j(t)),&\\
&{c}_{i} = \cos(\phi_i(t)), \qquad{c}_{i,j} =  \cos(\phi_i(t) - \phi_j(t)).&\\
&{sc}_{n} =  \frac{\sum_{j=1}^{n-1} c_j}{n^2} + \frac{c_n}{2n^2},\qquad
{ss}_{n} = \frac{\sum_{j=1}^{n-1} s_j}{n^2} + \frac{s_n}{2n^2}.&
\end{align}
\end{subequations}

The matrices in Eq.~(\ref{eq:HenckeyEquationsMatrixForm}) are constructed as follows:
\begin{enumerate}
    \item  Mass matrix $\mathbf{M}$: This is an $n\times n$ symmetric matrix ($\mathbf{M} =\mathbf{M}^\top$) with elements given by:
\begin{eqnarray}\label{eq:MassMatrix}
\mathbf{M} =& \nonumber \\ &
\resizebox{0.8 \textwidth}{!}{%
$\begin{pmatrix}
\frac{6n-7}{6n^3}  & \frac{(n - 1) {c}_{1,2}}{n^3}  & \cdots & \frac{3 {c}_{1,n-3}}{n^3} & \frac{2 {c}_{1,n-2}}{n^3} & \frac{{c}_{1,n-1}}{n^3} & \frac{{c}_{1,n}}{8n^3} \\
\frac{(n - 1) {c}_{1,2}}{n^3}  & \frac{6n-13}{6n^3}  & \cdots & \frac{3 {c}_{2,n-3}}{n^3} & \frac{2 {c}_{2,n-2}}{n^3} & \frac{{c}_{2,n-1}}{n^3} & \frac{{c}_{2,n}}{8n^3} \\
\vdots  & \vdots & \ddots & \vdots  & \vdots  & \vdots & \vdots  \\
\frac{3 {c}_{1,n-3}}{n^3}  & \frac{3 {c}_{2,n-3}}{n^3}  & \cdots & \frac{17}{6n^3} & \frac{2 {c}_{n-3,n-2}}{n^3} & \frac{{c}_{n-3,n-1}}{n^3} & \frac{{c}_{n-3,n}}{8n^3} \\
\frac{2 {c}_{1,n-2}}{n^3}  & \frac{2 {c}_{2,n-2}}{n^3}  & \cdots & \frac{2 {c}_{n-3,n-2}}{n^3} & \frac{11}{6n^3} & \frac{{c}_{n-2,n-1}}{n^3} & \frac{{c}_{n-2,n}}{8n^3} \\
\frac{{c}_{1,n-1}}{n^3}  & \frac{{c}_{2,n-1}}{n^3}  & \cdots & \frac{{c}_{n-3,n-1}}{n^3} & \frac{{c}_{n-2,n-1}}{n^3} & \frac{5}{6n^3} & \frac{{c}_{n-1,n}}{8n^3} \\
\frac{{c}_{1,n}}{8n^3}  & \frac{{c}_{2,n}}{8n^3}  & \cdots & \frac{{c}_{n-3,n}}{8n^3} & \frac{{c}_{n-2,n}}{8n^3} & \frac{{c}_{n-1,n}}{8n^3} & \frac{1}{24n^3} \\
\end{pmatrix}.$
} & \nonumber \\
\end{eqnarray}
\item Linear stiffness matrix $\mathbf{K}$: This is an $n\times n$ symmetric matrix ($\mathbf{K} =\mathbf{K}^\top$) with elements as follows:
\begin{equation}\label{eq:LinearStiffnessMatrix}
\mathbf{K} = n
\begin{pmatrix}
2  &-1  &0& \cdots &0 &0&0 \\
-1 &2  &-1& \cdots &0 &0&0 \\
0 &-1  &2& \cdots &0 &0&0 \\
\vdots  & \vdots & \vdots & \ddots &  \vdots & \vdots  & \vdots   
\\0  &0  & 0 & \cdots &2 &-1 &0 \\

0 &0  &0  & \cdots &-1 &2&-1 \\
0 &0  &0  & \cdots &0 &-1&2 
\end{pmatrix}.
\end{equation}
\item Nonlinear matrix $\mathbf{N}$: This is an $n\times n$ skew-symmetric matrix ($ \mathbf{N}=-\mathbf{N}^\top$). It represents a portion of the system’s geometric nonlinearity.

\begin{eqnarray}\label{eq:NonlinearStiffnessMatrix}
\mathbf{N} = \nonumber \\ &
\resizebox{0.8 \linewidth}{!}{%
$\begin{pmatrix}
0  & \frac{(n - 1) {s}_{1,2}}{n^3}  & \cdots & \frac{3 {s}_{1,n-3}}{n^3} & \frac{2 {s}_{1,n-2}}{n^3} & \frac{{s}_{1,n-1}}{n^3} & \frac{{s}_{1,n}}{8n^3} \\
-\frac{(n - 1) {s}_{1,2}}{n^3}  & 0  & \cdots & \frac{3 {s}_{2,n-3}}{n^3} & \frac{2 {s}_{2,n-2}}{n^3} & \frac{{s}_{2,n-1}}{n^3} & \frac{{s}_{2,n}}{8n^3} \\
\vdots  & \vdots & \ddots & \vdots  & \vdots  & \vdots & \vdots  \\
-\frac{3 {s}_{1,n-3}}{n^3}  & -\frac{3 {s}_{2,n-3}}{n^3}  & \cdots & 0 & \frac{2 {s}_{n-3,n-2}}{n^3} & \frac{{s}_{n-3,n-1}}{n^3} & \frac{{s}_{n-3,n}}{8n^3} \\
-\frac{2 {s}_{1,n-2}}{n^3}  & -\frac{2 {s}_{2,n-2}}{n^3}  & \cdots & -\frac{2 {s}_{n-3,n-2}}{n^3} & 0 & \frac{{s}_{n-2,n-1}}{n^3} & \frac{{s}_{n-2,n}}{8n^3} \\
-\frac{{s}_{1,n-1}}{n^3}  & -\frac{{s}_{2,n-1}}{n^3}  & \cdots & -\frac{{s}_{n-3,n-1}}{n^3} & -\frac{{s}_{n-2,n-1}}{n^3} & 0 & \frac{{s}_{n-1,n}}{8n^3} \\
-\frac{{s}_{1,n}}{8n^3}  & -\frac{{s}_{2,n}}{8n^3}  & \cdots & -\frac{{s}_{n-3,n}}{8n^3} & -\frac{{s}_{n-2,n}}{8n^3} & -\frac{{s}_{n-1,n}}{8n^3} & 0 \\
\end{pmatrix}.$
} & \nonumber \\
\end{eqnarray}

\item Stiffness matrix at the boundary $\mathbf{K}_{B}$: This matrix is obtained by summing the products of the total horizontal, vertical, and angular stiffness matrices with their respective stiffness coefficients (${K}_{x}^{Total}$, ${K}_{y}$, ${K}_{\phi}$):
\begin{equation}\label{eq:BoundaryStiffnessMatrix}
\mathbf{K}_{B}={k}_{x}^{Total} \mathbf{K}_{x} + {k}_y \mathbf{K}_{y}+ {k}_{\phi}\mathbf{K}_{\phi}
\end{equation}
The right boundary is physically modeled by three springs: a horizontal spring (${k}_{x}$), a vertical spring (${k}_{y}$), and an angular spring (${k}_{\phi}$), therefore:

\begin{equation}\label{eq:VirticalStiffnessMatrix}
\mathbf{K}_{y} = {ss}_{n}
\begin{pmatrix}
 c_1   &0 & \cdots &0 &0 \\
 0  & c_2  &\cdots &0 &0 \\
\vdots  & \vdots  & \ddots &  \vdots & \vdots    
\\

0 &0   & \cdots &c_{n-1} &0 \\
0 &0 & \cdots  &0&\frac{1}{2} c_n \\
\end{pmatrix},
\end{equation}
\begin{equation}\label{eq:rotationalStiffnessMatrix}
\mathbf{ K}_{\phi} = 
\begin{pmatrix}
0   &0 & \cdots &0 &0 \\
 0  & 0  &\cdots &0 &0 \\
\vdots  & \vdots  & \ddots &  \vdots & \vdots    
\\

0 &0   & \cdots &0 &0 \\
0 &0 & \cdots  &0&1 \\
\end{pmatrix}.
\end{equation}

An additional horizontal spring with stiffness ${k}_{St}$ is included to handle geometric nonlinearity due to stretching. This spring is connected in series with the horizontal spring at the right boundary (see Eq.~(\ref{eq:TotalHorizontalStiffness})). Hence, the total horizontal Stiffness Matrix $\mathbf{K}_{x}$ incorporates both the stretching effects and the stiffness of the horizontal boundary conditions:
\begin{equation}\label{eq:HorizontalStiffnessMatrix}
\mathbf{K}_{x} = \left(\frac{2n-1}{2n^2} - {sc}_{n}\right)
\begin{pmatrix}
 s_1   &0 & \cdots &0 &0 \\
 0  & s_2  &\cdots &0 &0 \\
\vdots  & \vdots  & \ddots &  \vdots & \vdots    
\\

0 &0   & \cdots &s_{n-1} &0 \\
0 &0 & \cdots  &0&\frac{1}{2} s_n \\
\end{pmatrix}.
\end{equation}
\item External force vector $\mathbf{F}$: This is an $n\times 1$ column vector representing the total force acting on each rigid link in a system:

\begin{eqnarray}\label{eq:ExcitationForceStiffnessMatrix}
&\mathbf{F} = F\cos(\Omega t)
\begin{pmatrix}
\frac{(n-1)c_1 }{n^2}   \\
 \frac{(n-2)c_2 }{n^2}   \\
\vdots \\

\frac{c_{n-1} }{n^2} \\
\frac{c_n}{8n^2} \\
\end{pmatrix}.&
\end{eqnarray}
\end{enumerate}

 This force arises due to the base excitation of the vertical force $F$. 
It's worth noting that the internal damping matrix $\mathbf{C}$ is related to the damping coefficient $d$ and the stiffness matrix, as could be realized from Eq.~(\ref{eq:HenckeyEquationsMatrixForm}) $\mathbf{D}=d\hspace{0.1cm}\mathbf{K}$.

For the sake of coding convenience, we have formalized these matrices in~\ref{appendix:A}.

Now, we employed {\rmfamily \scshape Wolfram Mathematica 13} to derive the ordinary differential equations (ODEs) utilizing Eq.~(\ref{eq:lagrange}) or Eq.~(\ref{eq:HenckeyEquationsMatrixForm}). These ODEs were then solved using {\rmfamily \scshape Matlab R2024a} (ode15s, which is specifically designed for stiff ODEs).
\section{Validation} \label{Validation}

For validation, we compared our obtained results with established findings from previous works \cite{rezaei2017two,Zamanian2015discretization,NIMAMAHMOODI2007577}, revealing a notable agreement. To facilitate the replication of experiments by other researchers, we selected a sample system similar to a widely available steel ruler beam, ensuring accessibility for future experimental investigations. Detailed properties and geometry of the system are presented in Table~\ref{Ta:properties}.

\begin{table}[H]
\caption{Geometrical and mechanical properties of the considered system}
\label{Ta:properties}
\resizebox{\linewidth}{!}{
\begin{tabular}{ccccccc}
\\\hline
$l$ (cm) & $w$ (cm) & $t_{b}$ (cm) & $E$ (GPa) & $\rho_{b}$ ({kg/m}$^3$) & $d$ \\ \hline
$50$ & $2.6$ & $0.8$ & $200$ & $7850$ & $0.002$ 
\\\hline
\end{tabular}}
\end{table}
In addition to the system properties, we have used several abbreviations to describe different aspects of deflection in beam models. These abbreviations are defined in table~\ref{Ta:abbreviations}.
\begin{table}[H]
\centering
\caption{Abbreviations}
\label{Ta:abbreviations}
\begin{tabular}{|l|p{0.6\linewidth}|}
\hline
\textbf{Abbreviation} & \textbf{Description} \\ \hline
LEB & Linear Euler-Bernoulli model discretized via Galerkin method \\
NEB & Nonlinear Euler-Bernoulli model discretized via Galerkin method \\
NH & Nonlinear Hencky beam Model method \\ \hline
\end{tabular}
\end{table}

The Linear Euler-Bernoulli (LEB) model, discretized via the Galerkin method described in subsection~\ref{Galerkinmethod}, provides a linear approximation of beam deflection by linearizing of Eq.~(\ref{eq:ShorteningWEquations}) or Eq.~(\ref{eq:StretchingOnedimensionalEquations}) around initial positions, eliminating nonlinear terms. The Nonlinear Euler-Bernoulli (NEB) model, also discretized using the Galerkin method, accounts for small deflections and geometric nonlinearity in the equations of motion from Eqs.~(\ref{eq:ShorteningWEquations}) and~(\ref{eq:StretchingOnedimensionalEquations}). The Nonlinear Hencky (NH) beam Model employs Hencky theory for analyzing large deformations and geometric nonlinearity in beams, as detailed in section~\ref{sec:Hencky}.

To achieve convergence in our calculations, we first employed $25$ elements $n=25$ within the Nonlinear Hencky’s beam Models and utilized 3 modes $N_m=3$ when applying the Galerkin method. To prevent overwhelming readers with too many figures, we presented C-F and C-C beams, serving as examples of systems with shortened and stretched boundaries, respectively. We set $k_x$, $k_y$, and $k_\phi$ to $10^8$, making them effectively large enough to be considered as infinity for C-C cases.

Due to the challenges of the Euler-Bernoulli beam theory, the Galerkin method applied within this framework can only accurately predict small beam deflections. Therefore, we pick the excitation force strengths ($F$) to keep the beam deflections small, as the Euler-Bernoulli theory requires.
\begin{figure}[H]
\centering
       \begin{subfigure}{\textwidth} \centering
        \includegraphics[width=0.7  \linewidth]{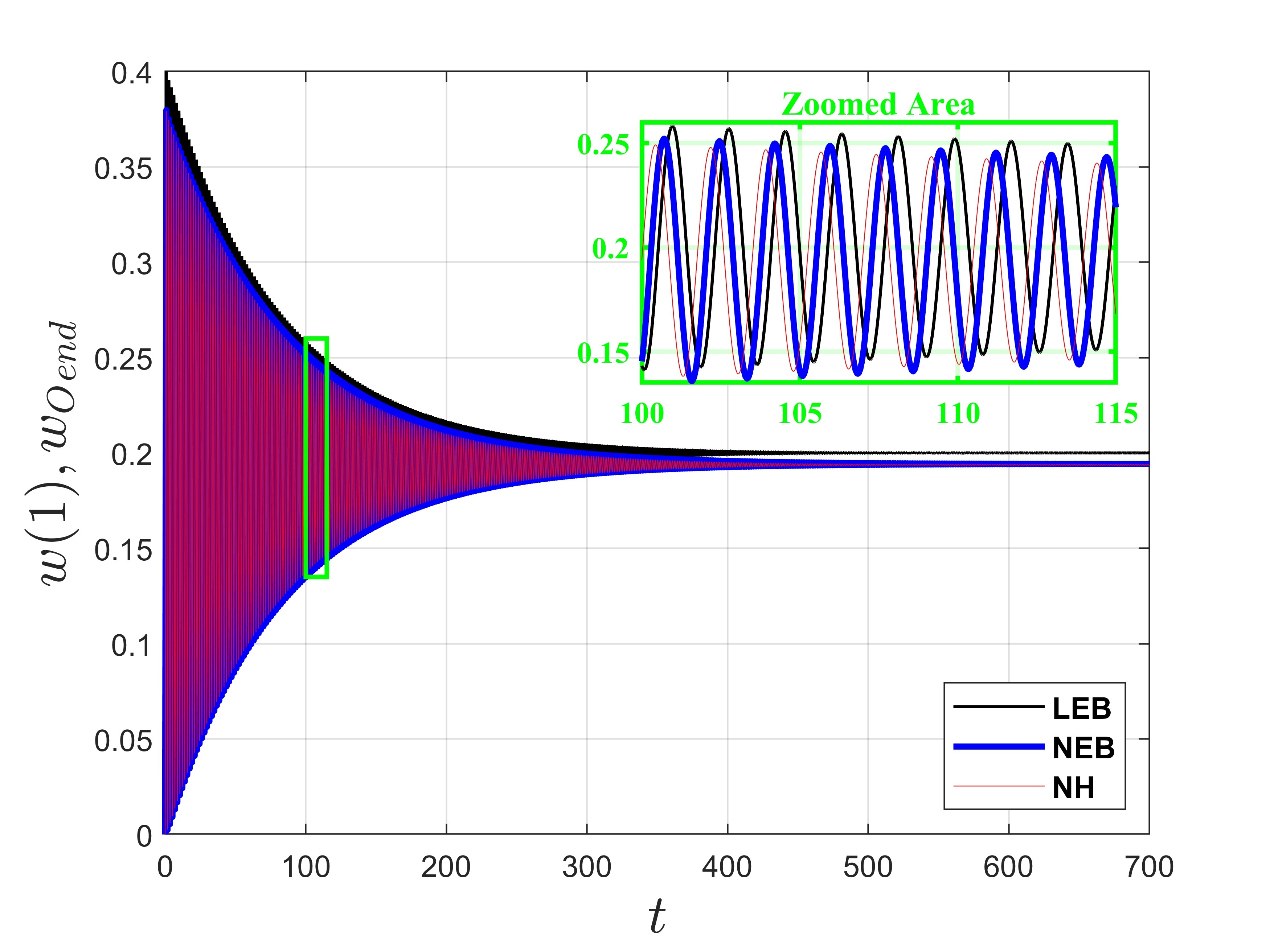}
        \subcaption{}
        \label{figure5-a}
    \end{subfigure}
       \begin{subfigure}{\textwidth} \centering
        \includegraphics[width= 0.7 \linewidth]{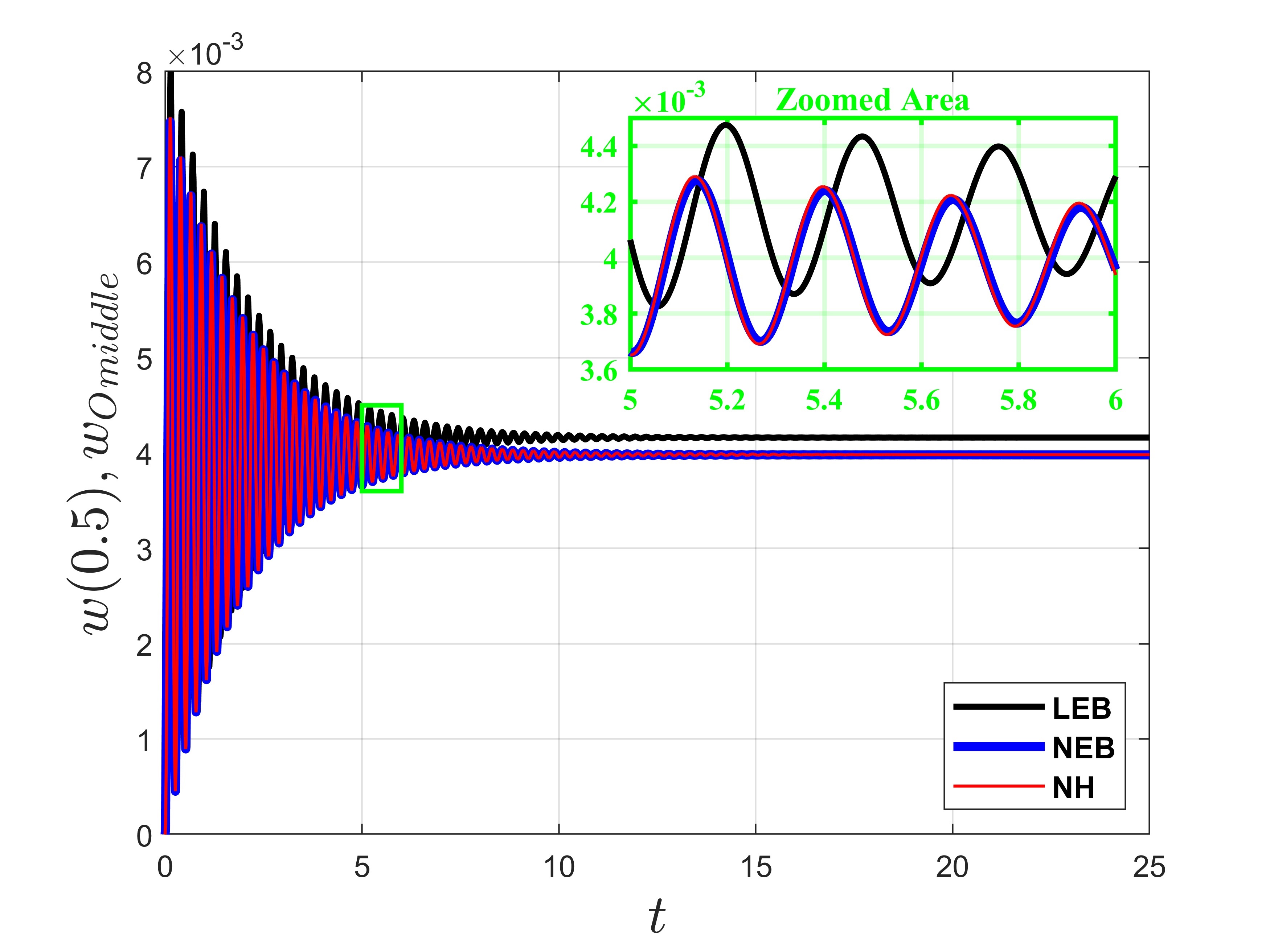}
        \subcaption{}
        \label{figure5-b}
    \end{subfigure}
       \caption{Comparison of NH with 26 elements ($n=25$), LEB, and NEB with 3 modes ($N_m=3$): Static time history analysis ($\Omega=0$) subjected to a force of $F=1.6$ \textbf{(a)} C-F beam (midpoint). \textbf{(b)} C-C beam (midpoint). \label{figure5}}
\end{figure}
As observed from Fig.~\ref{figure5}, NEB and NH match for both C-F and C-C beams under static conditions with a force of $F=1.6$, all being within the range of small deflections. However, there is a slight variation between LEB and the two nonlinear models (NEB and NH), due to LEB considering only linear parts.

\begin{figure}[H]
\centering
       \begin{subfigure}{\textwidth} \centering
        \includegraphics[width=0.7 \linewidth]{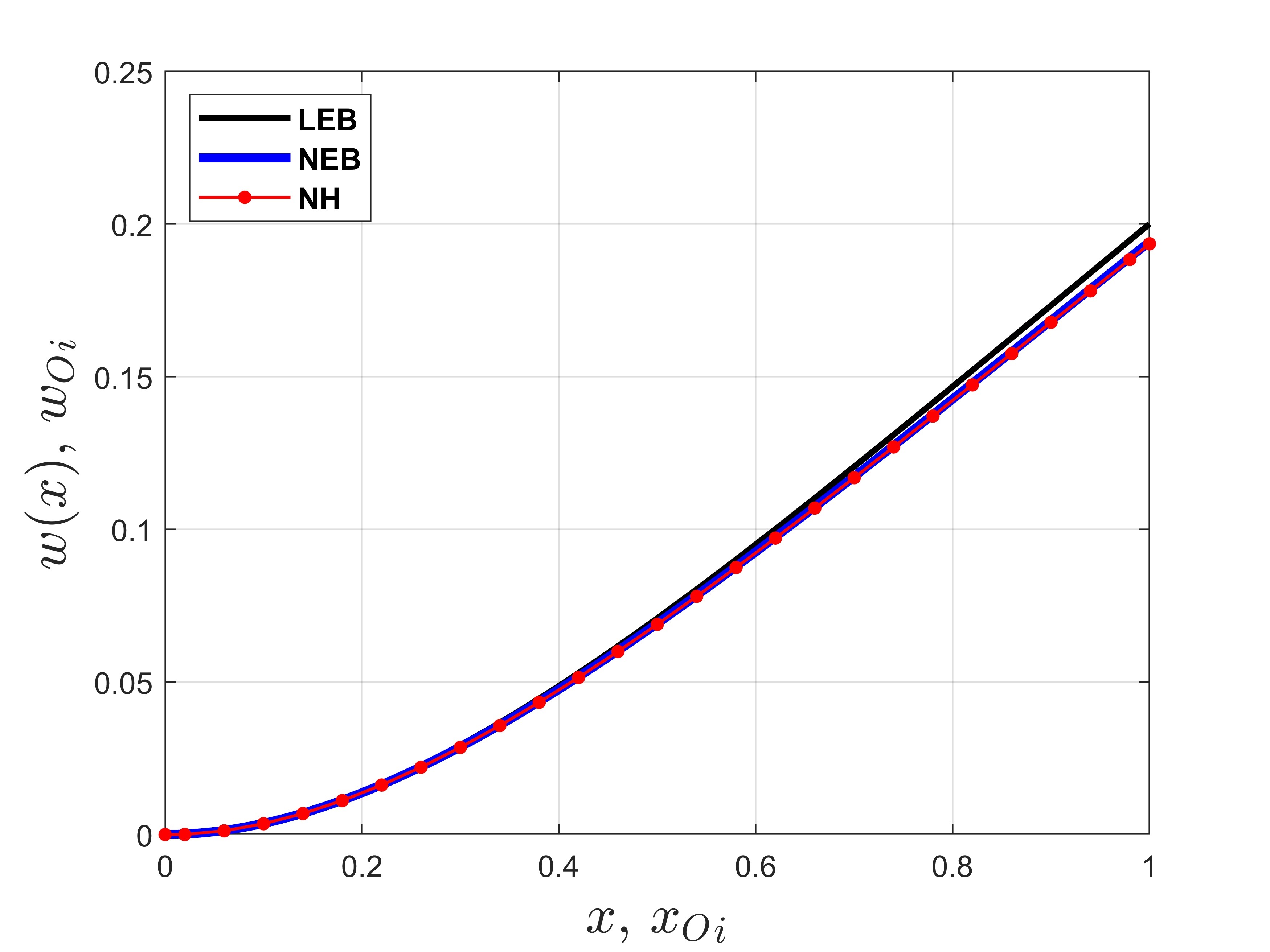}
        \subcaption{}
        \label{figure6-a}
    \end{subfigure}
       \begin{subfigure}{\textwidth} \centering
        \includegraphics[width=0.7 \linewidth]{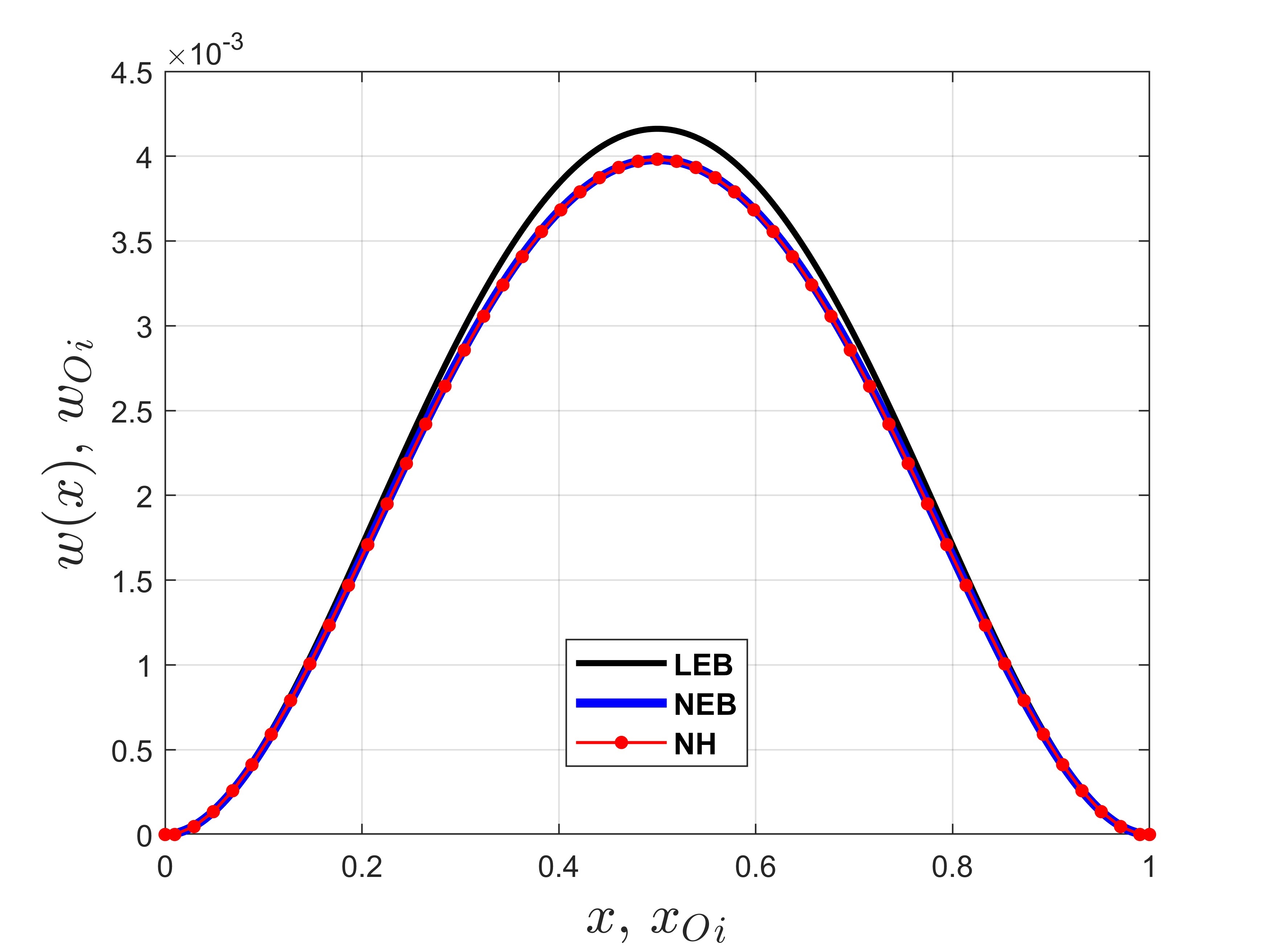}
        \subcaption{}
        \label{figure6-b}
    \end{subfigure}
       \caption{Comparison of NH with 26 elements ($n=25$), LEB, and NEB with 3 modes ($N_m=3$): Static deflection ($\Omega=0$) after stabilization subjected to a force of $F=1.6$. \textbf{(a)} C-F beam at $t=700$. \textbf{(b)} C-C beam at $t=25$. \label{figure6}}
\end{figure}

We use subscripts $_{middle}$ and $_{end}$ to refer to the middle and end joints of the beam, respectively. In Fig.~\ref{figure6}, in the realm of linear static deflection, LEB displays slight variations along the beam when contrasted with NH. However, concerning both C-F and C-C configurations, NEB closely aligns with the outcomes of NH with just 25 elements. This similarity holds true throughout the entire beam.
\begin{figure}[H]
\centering
       \begin{subfigure}{\textwidth} \centering
\centering
        \includegraphics[width=0.7 \linewidth]{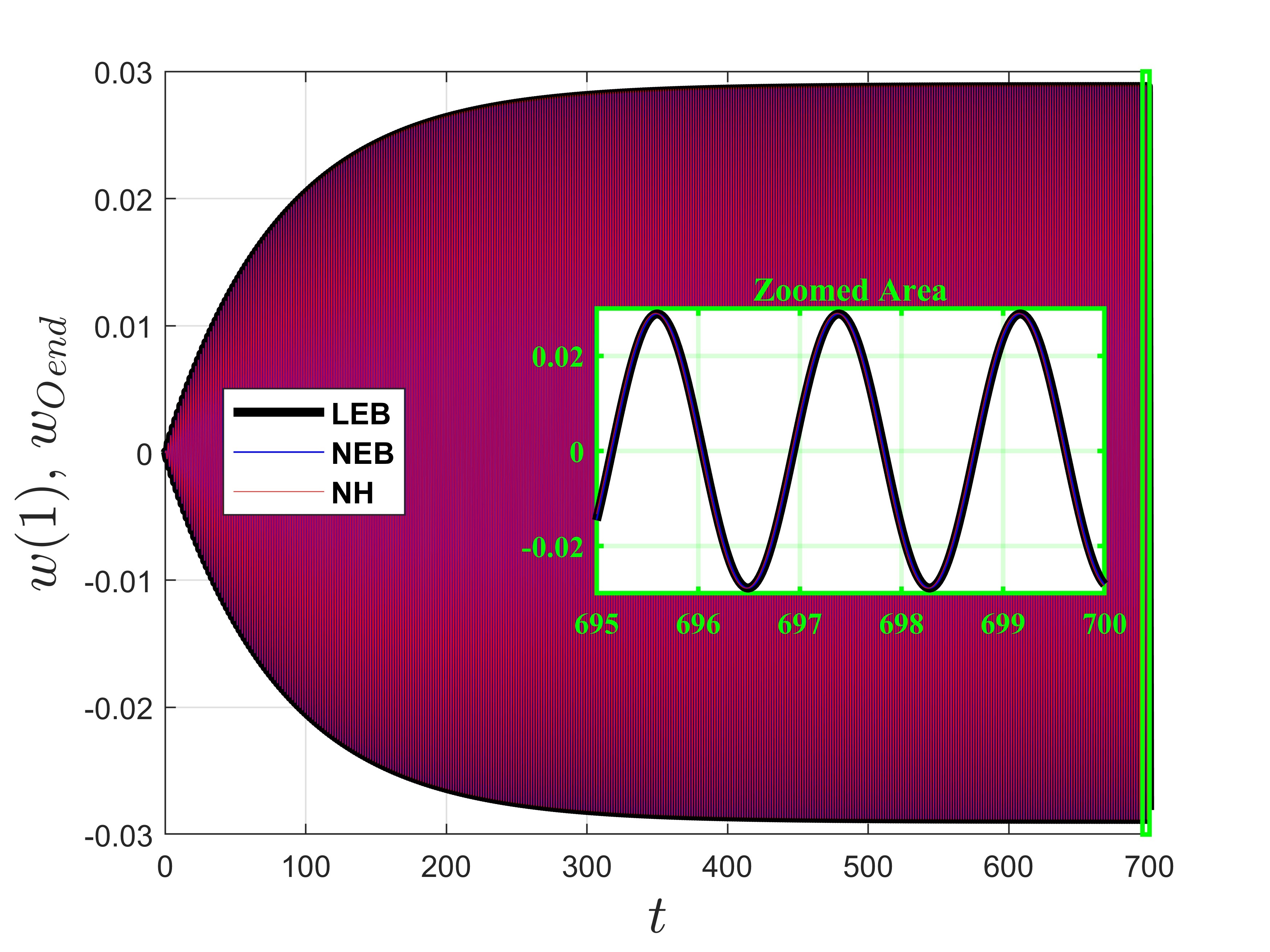}
       \subcaption{}
        \label{figure7-a}
    \end{subfigure}
       \begin{subfigure}{\textwidth} \centering
\centering
        \includegraphics[width=0.7 \linewidth]{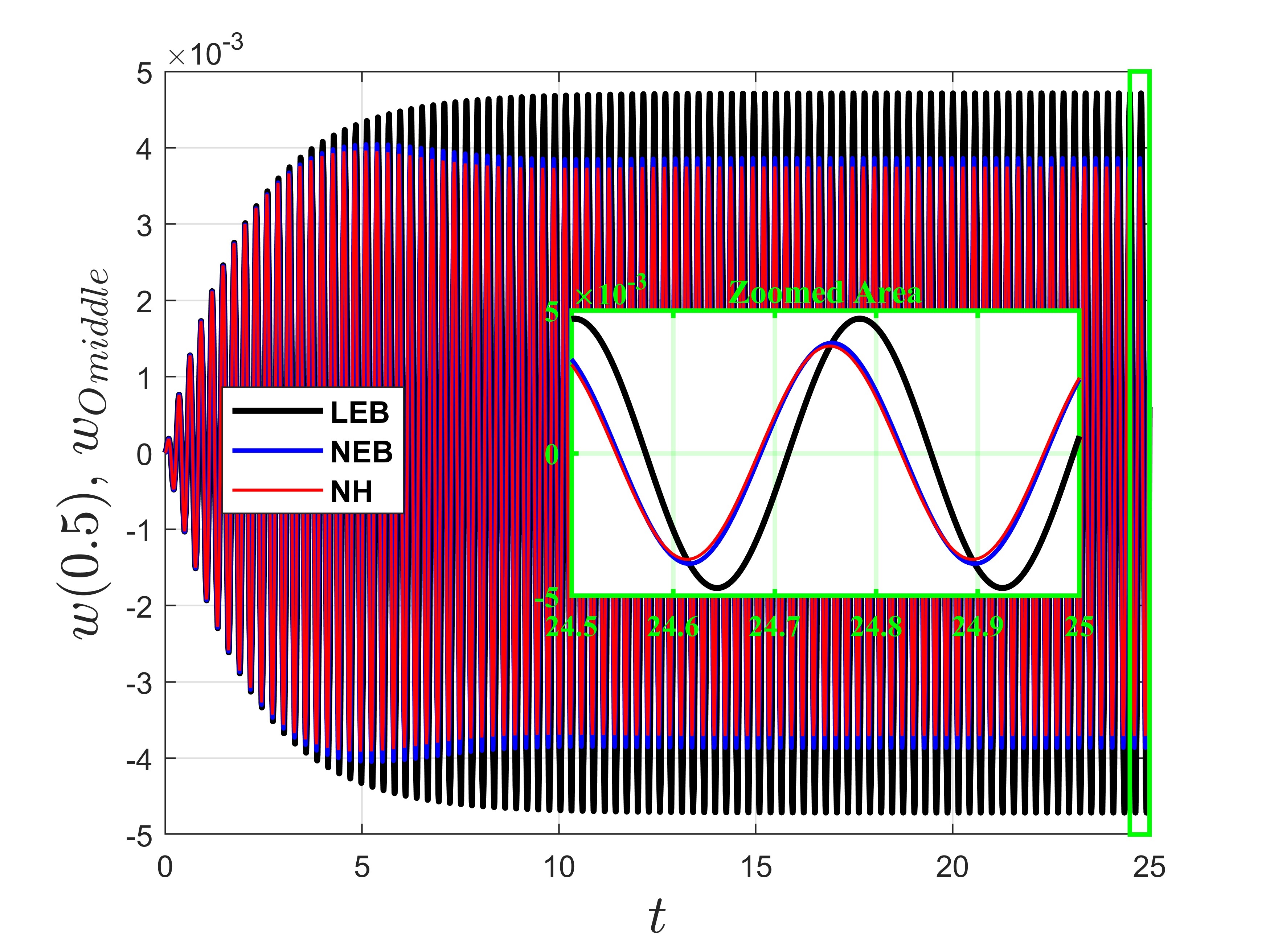}
 \subcaption{}
        \label{figure7-b}
    \end{subfigure}
       \caption{Comparison of NH with 26 elements ($n=25$), LEB, and NEB with 3 modes ($N_m=3$): Time History Analysis near the first resonance frequency $\Omega=\omega_1$. \textbf{(a)}, C-F beam (endpoint) subjected to a force of $F=0.0016$ from $t=0$ to $t=25$. \textbf{(b)} C-C beam (midpoint) subjected to a force of $F=0.08$ from $t=0$ to $t=700$. \label{figure7}}
\end{figure}

In the domain of dynamic analysis, comparing results at the first resonance frequency is essential. The first resonance frequencies are $\omega_1=3.52$ for the C-F beam and $\omega_1=22.37$ for the C-C beam. We used different excitation forces for C-F (Fig.~\ref{figure7-a}) and C-C cases (Fig.~\ref{figure7-b}) to experience small deflections for beams, small enough to justify Euler Bernoulli's theory. For the C-F beam, all results are the same. For C-C beams, there is a difference between linear and nonlinear solutions; however, NEB and the NH show good agreement (with an ignorable difference) using only$n=25$.
\begin{figure}[H]
\centering
\includegraphics[width= 0.7 \linewidth]{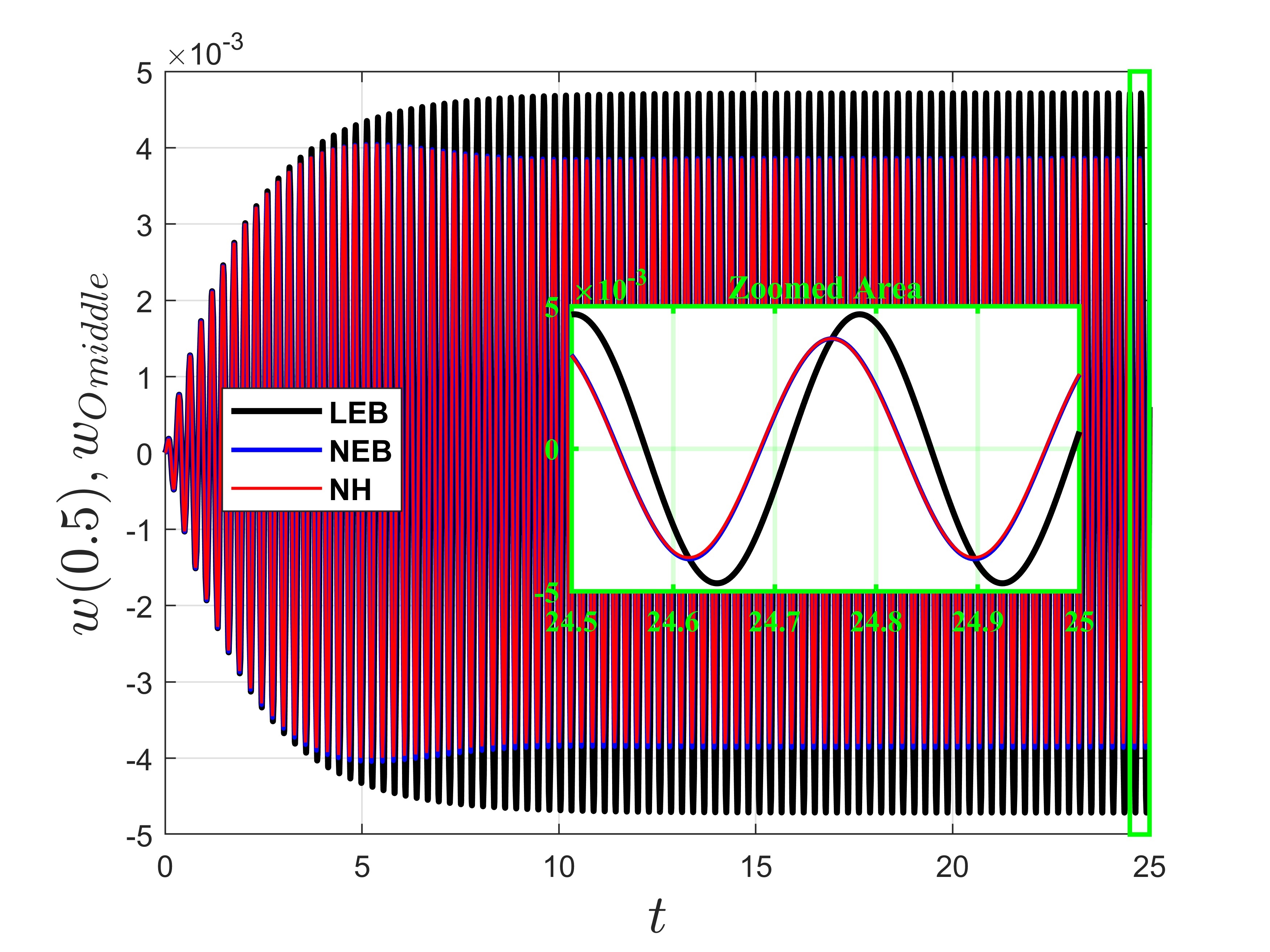}
 \caption{Comparison of NH with $52$ elements (n=51), LEB, and NEB with 3 modes ($N_m=3$): Time History Analysis near the first resonance frequency $\Omega=\omega_1$ for a C-C beam (midpoint) subjected to a force of $F=0.08$ from $t=0$ to $t=700$. \label{figure8}}
\end{figure} 
 We added another figure, Fig.~\ref{figure8}, with results for 51 elements (choosing 51 because the middle point of the beam, lies exactly in the $26th$ joint, $x=x_{O_{26}}=x_{O_{middle}}=0.5$). Hence, with increasing the number of elements $n$, the difference between NH and NEB has significantly reduced. 

\begin{figure}[H]
\centering
       \begin{subfigure}{\textwidth} \centering
        \includegraphics[width=0.7 \linewidth]{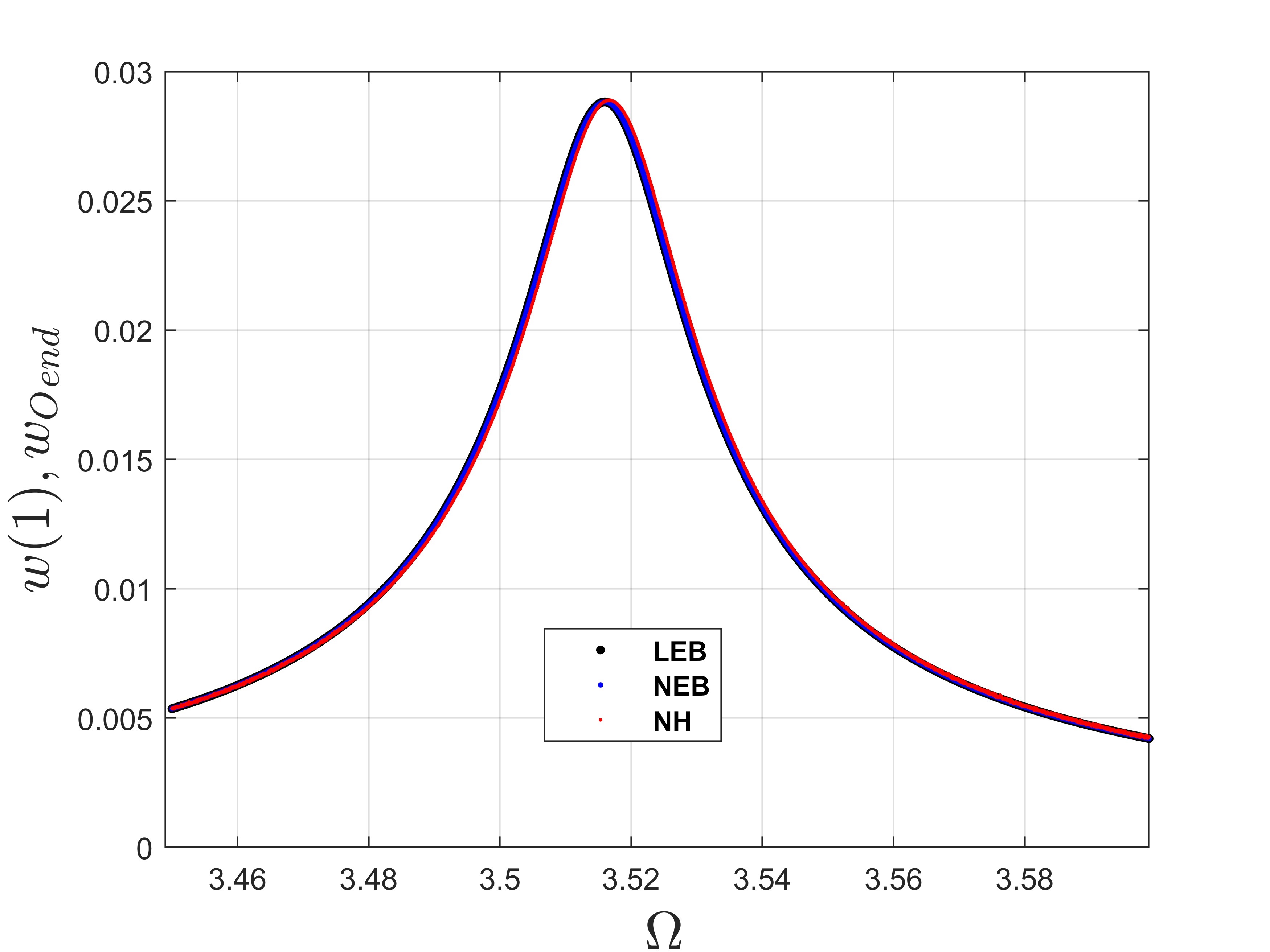}
        \subcaption{}
        \label{figure9-a}
    \end{subfigure}
       \begin{subfigure}{\textwidth} \centering
        \includegraphics[width=0.7 \linewidth]{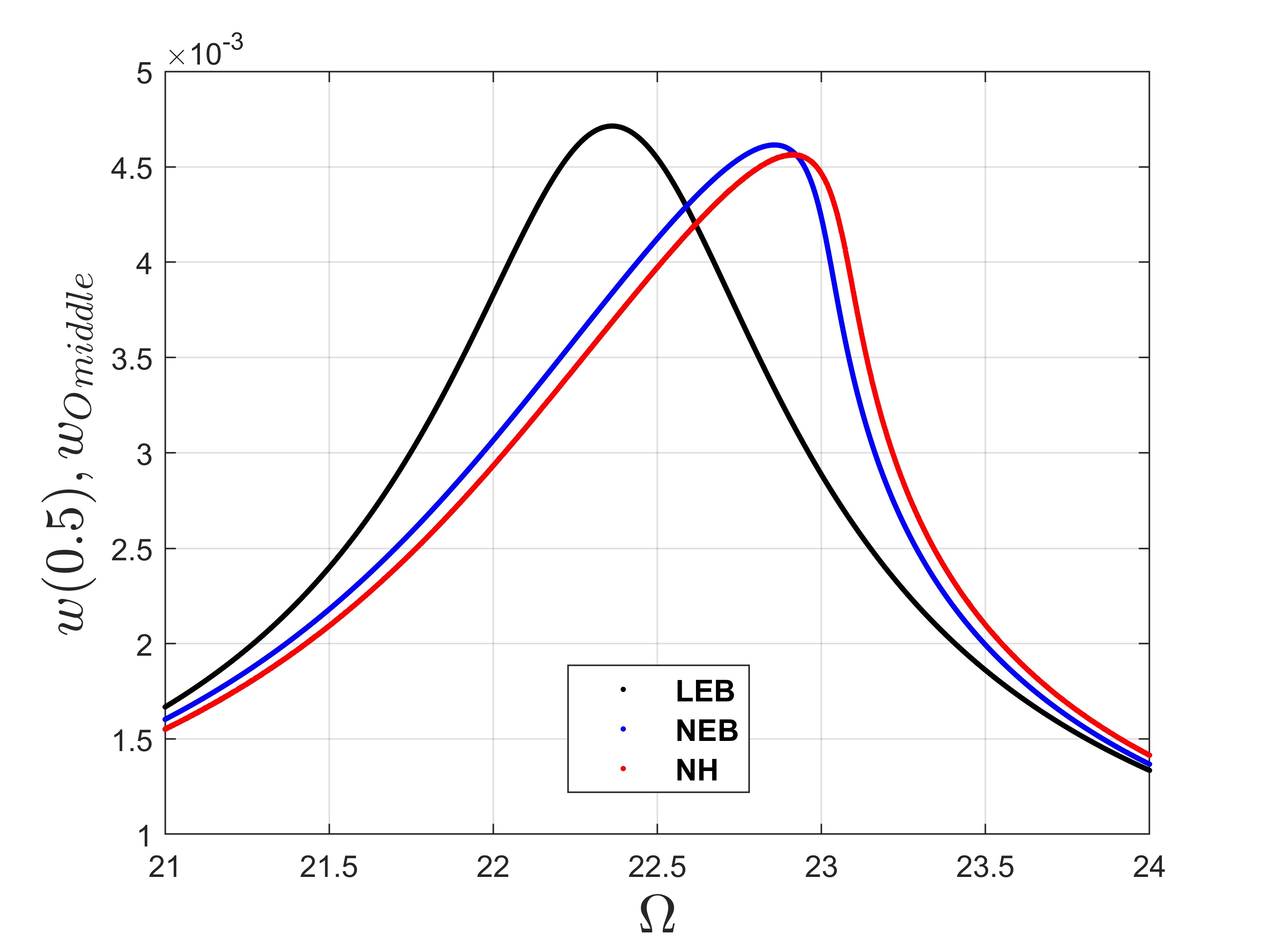}
        \subcaption{}
        \label{figure9-b}
    \end{subfigure}
       \caption{Comparison of NH with 26 elements ($n=25$), LEB, and NEB with 3 modes ($N_m=3$): Frequency Response Analysis near the first resonance frequency ($\Omega \approx \omega_1$). \textbf{(a)} C-F beam (endpoint) subjected to a force of $F=0.0016$. \textbf{(b)} C-C beam (midpoint) subjected to a force of $F=0.08$. \label{figure9}}
\end{figure}
\begin{figure}[H]
\centering
\includegraphics[width= 0.7 \linewidth]{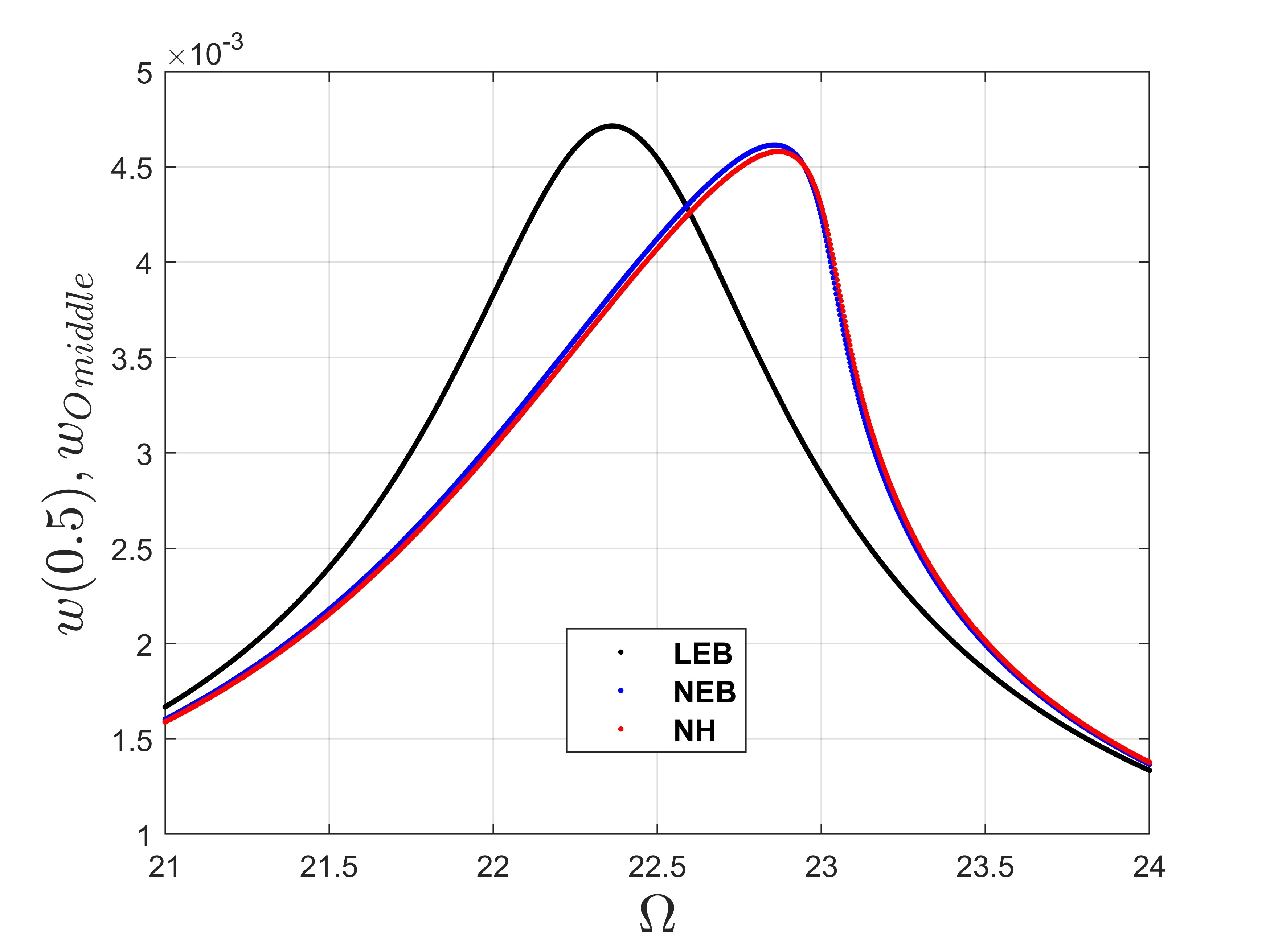}
 \caption{Comparison of NH with 52 elements $(n=51)$, LEB, and NEB with 3 modes ($N_m=3$): Frequency Response Analysis near the first resonance frequency ($\Omega \approx \omega_1$) for C-C beam (midpoint) subjected to a force of $F=0.08$. \label{figure10}}
\end{figure}

To assess how different methods handle nonlinearities, we present frequency resonance responses in Fig.~\ref{figure9}. While Euler-Bernoulli's theory effectively describes the behavior of structures experiencing small deflections, the difference between linear and nonlinear solutions is minimal for cantilever configurations, especially with the given dimensions. For the C-C beam, a difference is observed between the linear and nonlinear solutions. When NH is applied with $n=25$, it does not quite align with NEB. However, by increasing $n$ to $51$ in NH, the alignment improves significantly, as demonstrated in Fig.~\ref{figure10}.
This figure indicates that the gap between the nonlinear outcomes from NEB and NH has notably decreased by increasing the number of elements. Therefore, our validation demonstrates strong agreement between the nonlinear results obtained through Euler-Bernoulli beam models and Henckey's beam models across various static and dynamic analyses. As previously mentioned in section~\ref{sec:EulerBernoulli}, NEB depends on boundary conditions This allows for an interesting investigation of the nonlinearity of partly-shortened or partly-stretched beams, which may not be possible with the Galerkin method employing nonlinear Euler-Bernoulli theory (NEB). Such exploration will be conducted in the following section.

\section{Unique application of nonlinear Hencky's beam models}\label{Sec:UniqueApplication}

Despite their computational demands, NH models are valid for large deflections and provide valuable insights. This section discusses one of the applications of NH that is impossible with NEB. We here explore the nonlinear behavior of a non-classical support configuration shown in Fig.~\ref{figure3}. As noted in subsection~\ref{subsec:NeitherStretchedNorShortened}, the nonlinear analysis of these boundary conditions ($k_y=0$, $k_\phi=0$ and $0 < k_x < \infty$) is beyond the capabilities of NEB. In this range of $k_x$ beams are partly shortened or stretched. NH addresses this limitation, which we will examine in this section. 

\begin{figure}[H]
\centering
\includegraphics[width=1\linewidth]{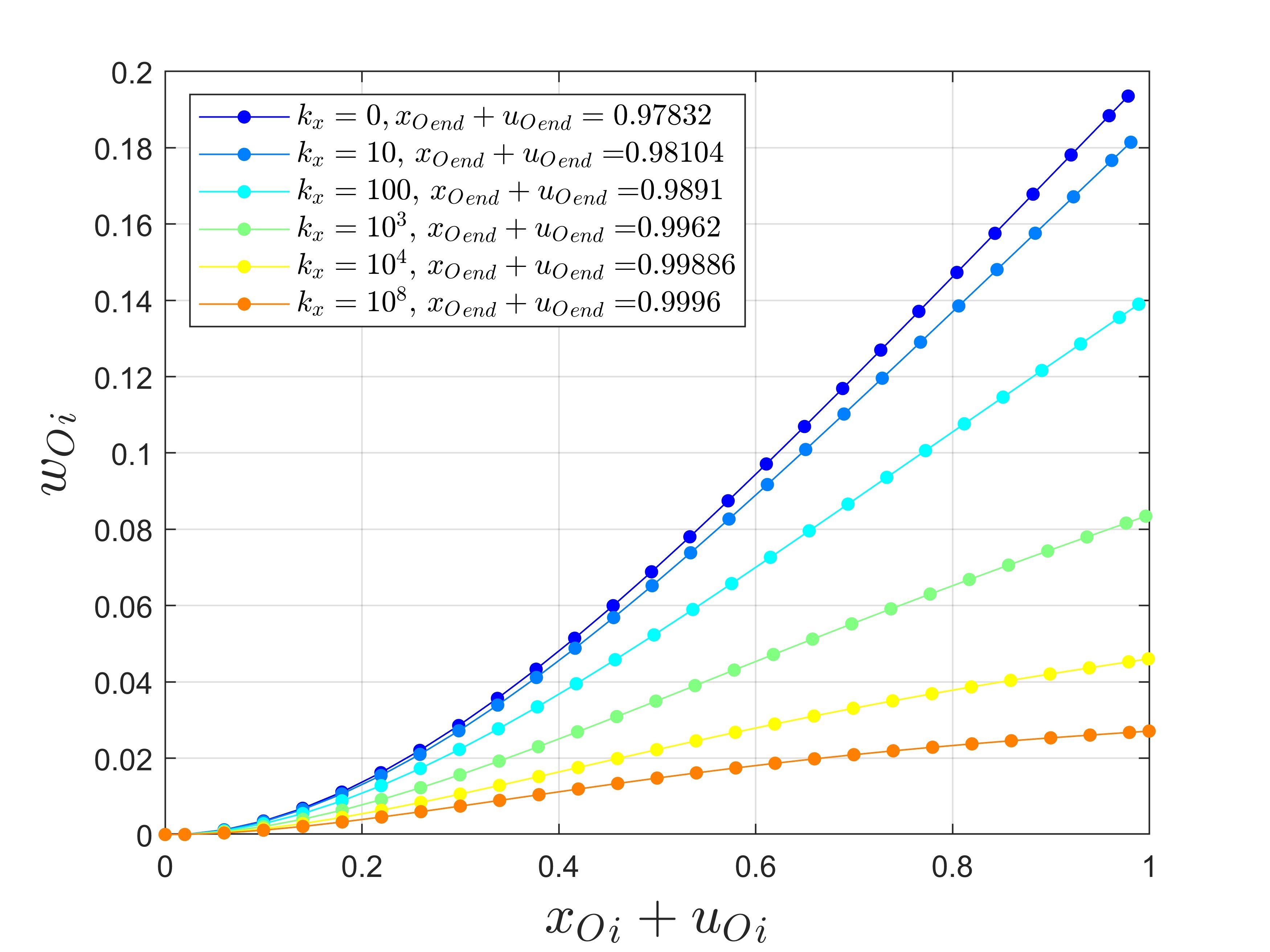}
\caption{Static deflections ($\Omega=0$) after stabilization ($t=700$) under a force of $F=1.6$ with varying horizontal spring stiffness ($k_x$), using NH ($n=25$).\label{figure11}}
\end{figure}
Fig.~\ref{figure11} shows the horizontal and vertical static displacements ($u$ and $w$) for various horizontal spring stiffnesses.
The graph shows that increasing horizontal spring stiffness results in lower deflection values. The peak deflection is observed at the beam’s end (${x_O}_{end}+{u_O}_{end}$) and tends toward $1$ as stiffness increases. Therefore, enhancing the spring's stiffness makes the system more rigid in the static region.

\begin{figure}[H]
\centering
\includegraphics[width=1\linewidth]{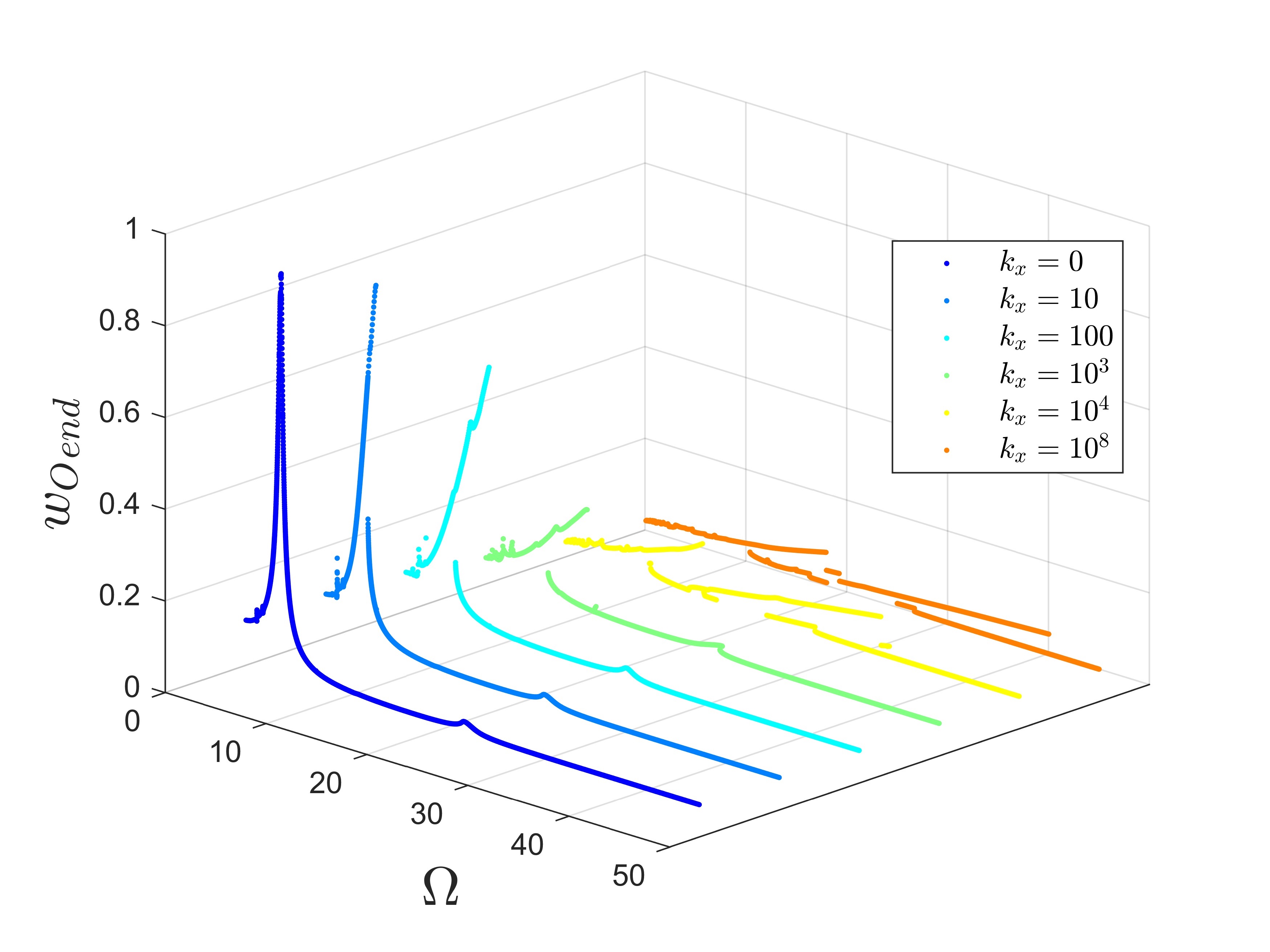}
\caption{Frequency responses ($0 \leq \Omega \leq 45$) after stabilization ($t=700$) under a force of $F=0.8$ with varying horizontal spring stiffness ($k_x$), using NH ($n=25$). \label{figure12}}
\end{figure}

We applied a very high force to create a large deflection. This allowed us to clearly see changes in the nonlinearity of the frequency response graph. The first resonance is always much larger than the second. To observe the nonlinear behavior in the second resonance, we needed a very high force, which also caused a large deflection at the first resonance. Fig.~\ref{figure12} shows a plot of the frequency response from zero to the second resonance frequency (for the cantilever $\omega_1=3.52$ and  $\omega_2=22.03$) for various horizontal spring stiffnesses. As stiffness increases, the peaks shift to the right, indicating that the first and second resonance frequencies increase. Thus, increasing horizontal stiffness makes the system more rigid dynamically.

\begin{figure}[H]
\centering
\includegraphics[width=1\linewidth]{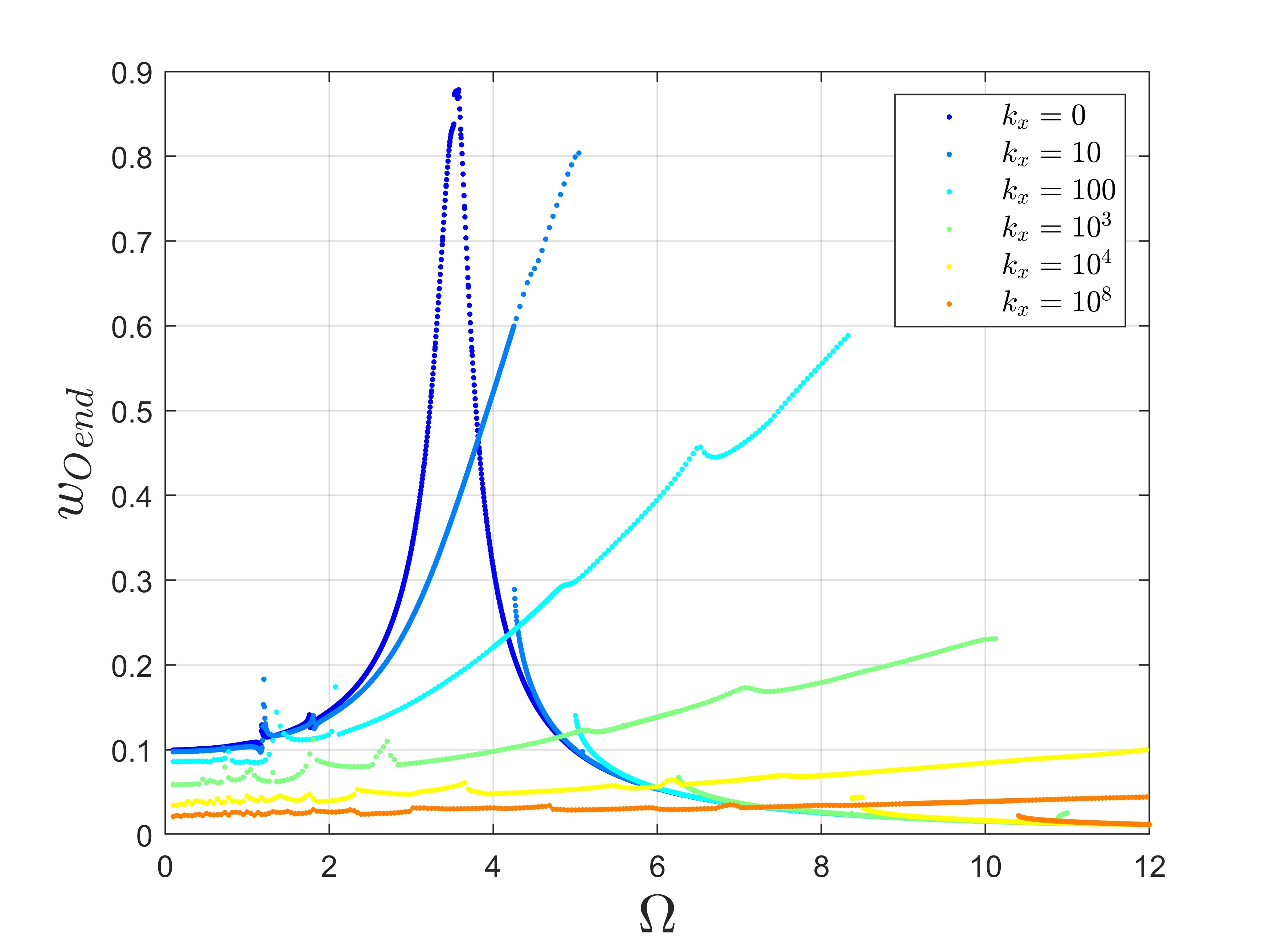}
\caption{Frequency responses near the first resonance frequency ($0 \leq \Omega \leq 10$) after stabilization ($t=700$) under a force of $F=0.8$ with varying horizontal spring stiffness ($k_x$), using NH ($n=25$).\label{figure13}}
\end{figure}

Fig.~\ref{figure13} shows frequency responses near the first resonance frequency. It reveals that the first resonant frequency and maximum deflection decrease with increasing stiffness, making the system stiffer. As illustrated, the nonlinearity becomes stronger near the first resonance by increasing  $k_x$, as shown by the steepening of the frequency response curves.

\begin{figure}[H]
\centering
\includegraphics[width=1\linewidth]{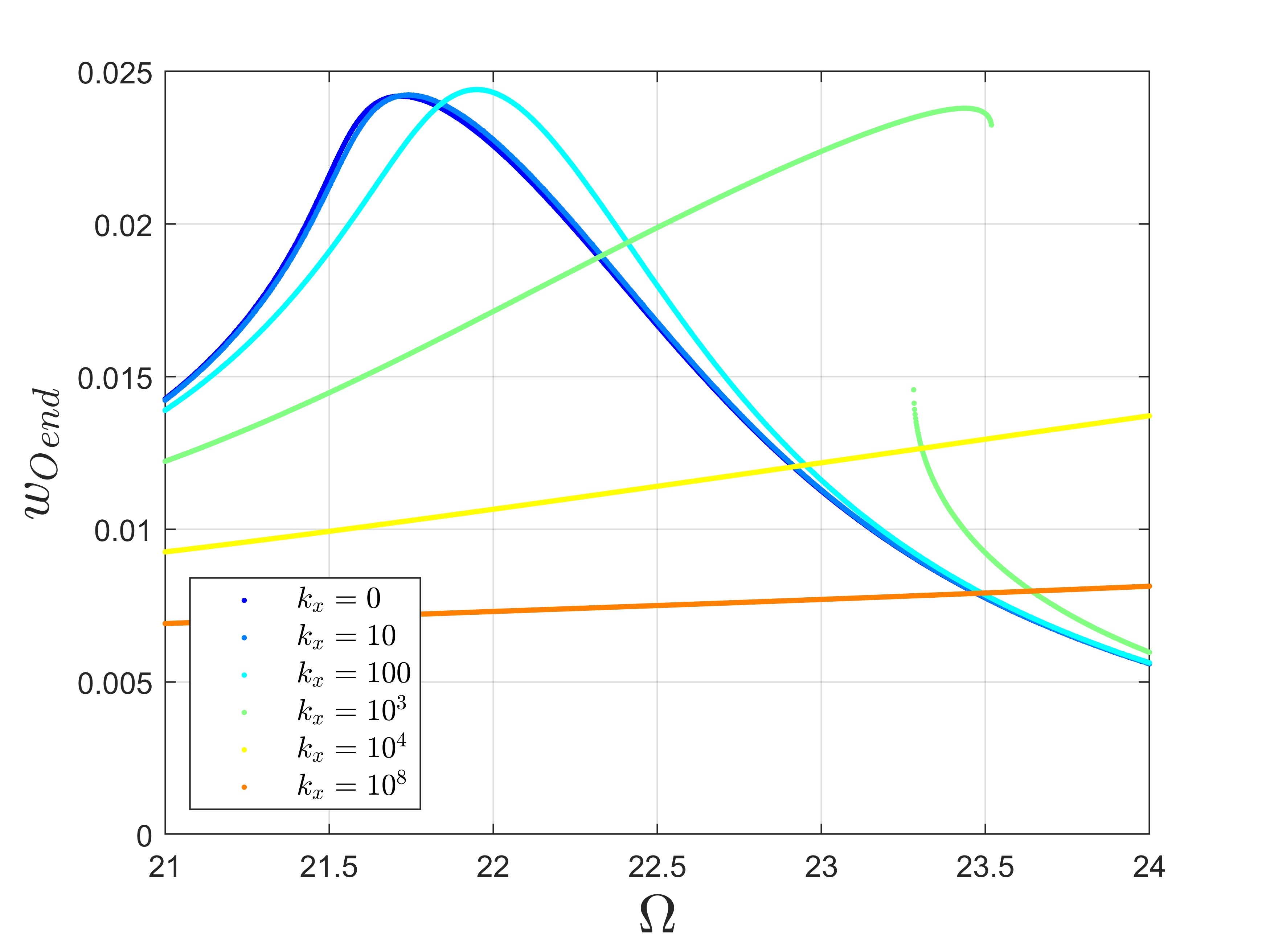}
 \caption{Frequency responses near the second resonance frequency ($21 \leq \Omega \leq 24$) after stabilization ($t=700$) under a force of $F=0.8$ with varying horizontal spring stiffness ($K_x$), using NH ($n=25$). \label{figure14}}
\end{figure}

\begin{figure}[H]
\centering
\includegraphics[width=0.75\linewidth]{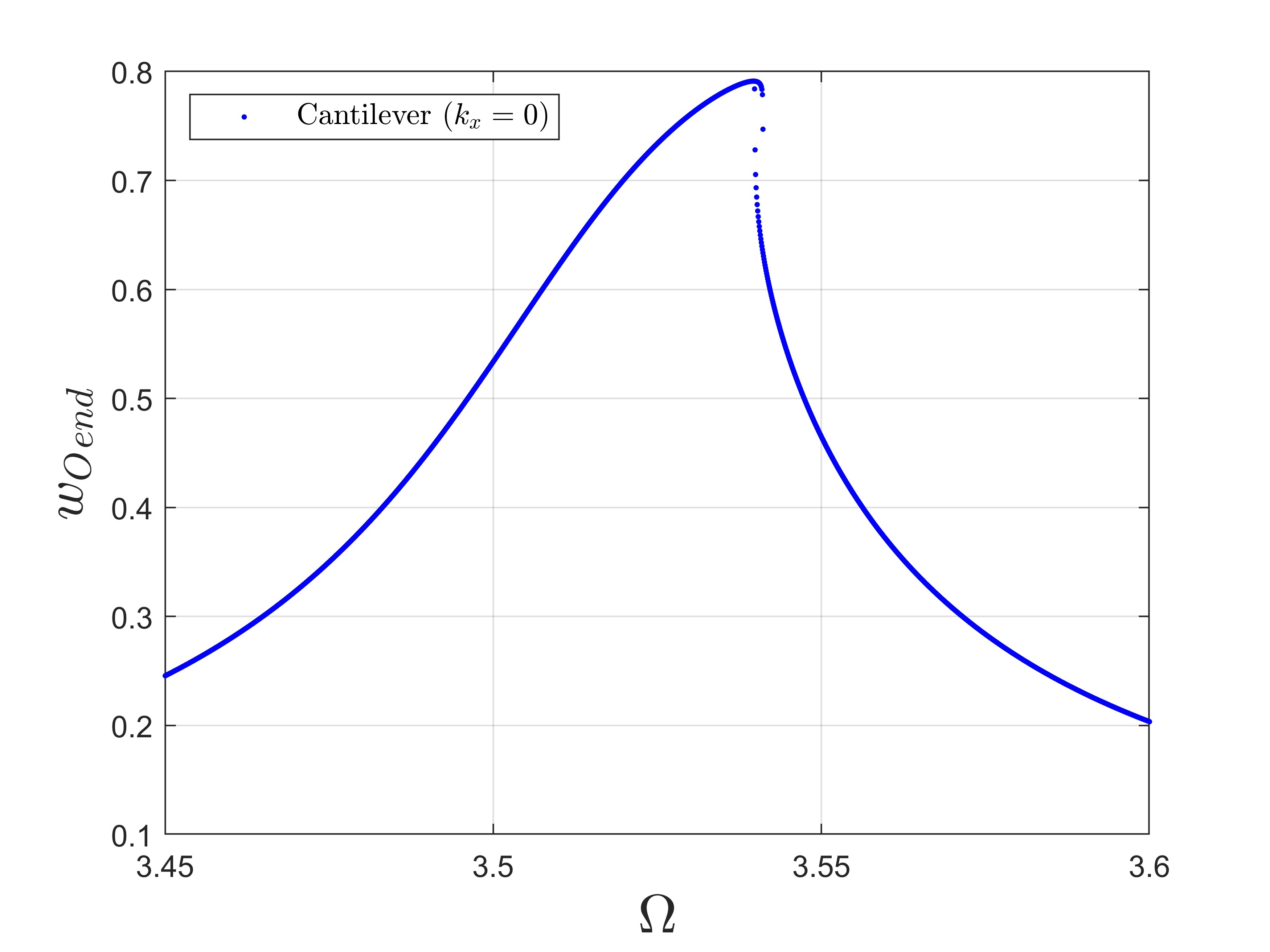}
\caption{Frequency responses near the first resonance frequency ($3.4 \leq \Omega \leq 3.6$) after stabilization ($t=700$) under a force of $F=0.08$ for a cantilever ($k_x=0$), using NH ($n=25$).}
\label{figure15}
\end{figure}

\begin{figure}[H]
\centering
\includegraphics[width=0.75\linewidth]{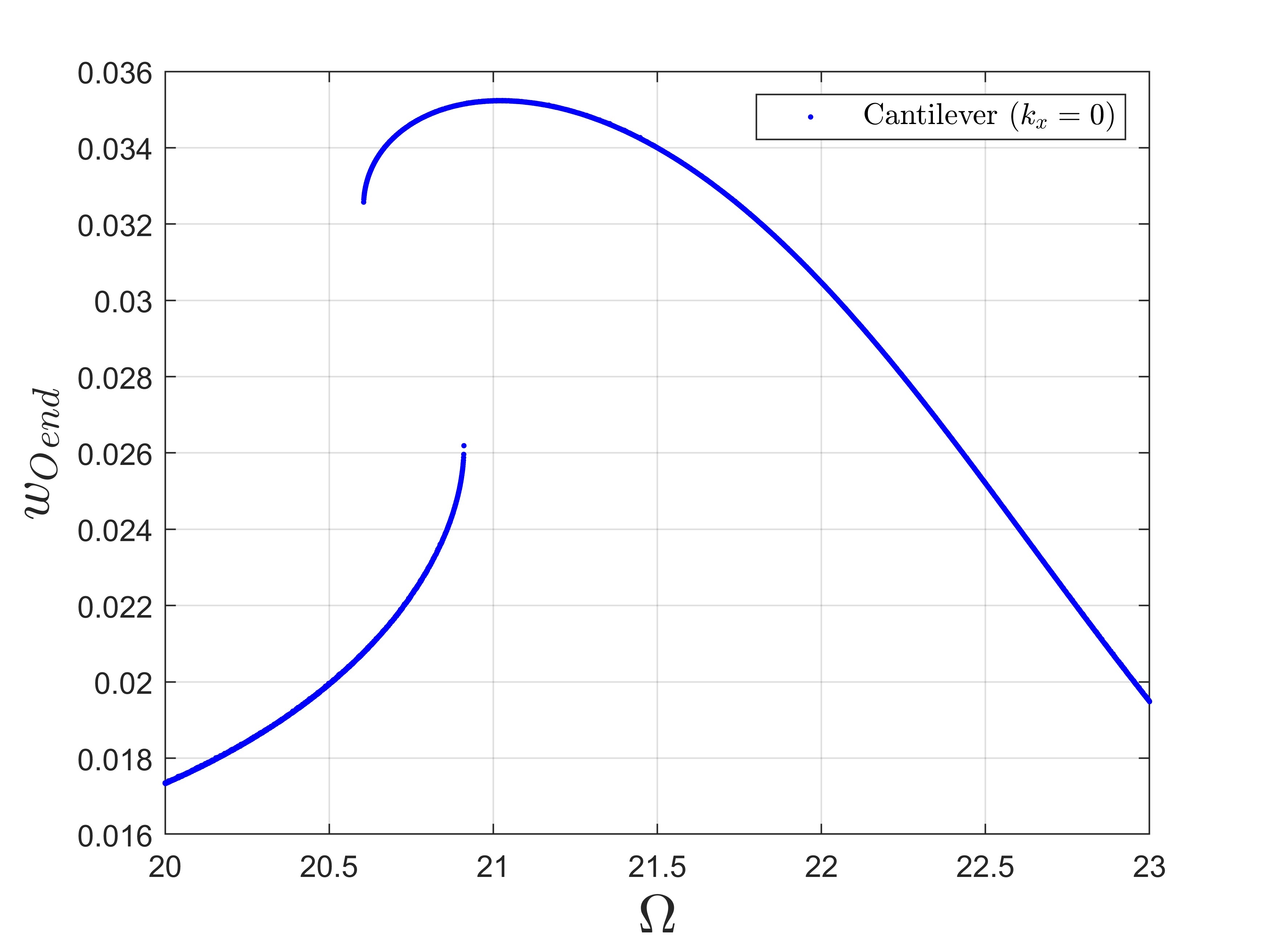}
\caption{Frequency responses near the second resonance frequency ($20 \leq \Omega \leq 23$) after stabilization ($t=700$) under a force of $F=1.6$ for a cantilever ($k_x=0$), using NH ($n=25$).}
\label{figure16}
\end{figure}

Fig.~\ref{figure14}, concerning the second resonance frequency, shows that as spring stiffness increases, the beam shows softening behavior and then hardening behavior, which can be explained by Eq.~\ref{eq:OvercomingStretchinShortening}. Before $k_x \leq 100$, the nonlinearity weakens (the beam behavior is partly softening), and afterward, it strengthens (the beam behavior becomes partly hardening). Around $k_x=100$, near the second resonance frequency, the system exhibits linear behavior, neither softening nor hardening behavior. This transition suggests a bifurcation point near the second resonance near $k_x=100$, where the system's behavior changes fundamentally (similar to \cite{Qiao2022}).
As observed from the cantilever figures, the dynamic behavior of cantilevers at their first and second resonance frequencies shows distinct differences. To illustrate this more clearly, we plotted the responses of the first ($F=0.08$) and second ($F=1.6$) resonance frequencies under varying excitation forces in Fig.~\ref{figure15} and Fig.~\ref{figure16}.

Interestingly, as shown in Fig.~\ref{figure15}, the cantilever exhibits hardening behavior near the first resonance frequency, while Fig.~\ref{figure16} illustrates softening behavior near the second resonance frequency. Although these observations might seem unexpected at first glance, Nayfeh and Pai \cite{Nayfeh1989} confirm these findings in their study of the nonlinear behavior of cantilever beams. They discovered that this nonlinear behavior arises from a combination of hardening and softening effects. The overall nonlinear response depends on the balance between these two effects. Specifically, the first resonance mode demonstrates hardening, whereas higher modes tend to exhibit softening \cite{doi:https://doi.org/10.1002/9783527617562.ch4}.

In conclusion, increasing the horizontal stiffness of the boundary condition makes the system stiffer. Although maximum deflection decreases near the first resonance frequency, it increases near the second resonance frequency. A bifurcation point near the second resonance frequency indicates that the system transitions from softening to hardening behavior.

\section{Conclusion} \label{Sec:Conclusion}
This paper addressed the challenges of the Euler-Bernoulli beam theory concerning assumptions about beam shortening and stretching. We explored boundary conditions, such as a cantilever with a horizontal spring, which causes beams to partly shorten or stretch based on spring stiffness. The traditional Euler-Bernoulli model may not accurately model these beams' geometric nonlinearity and is only suitable for small deflections.

To overcome these challenges, we presented nonlinear Hencky's beam models capable of describing partly stretched or shortened beams. Validation against the nonlinear Euler-Bernoulli model using the Galerkin method confirmed the effectiveness of Hencky's models for cantilever and clamped-clamped configurations, representing shortened and stretched beams.
We conducted a nonlinear analysis of a cantilever with a horizontal spring with varying stiffness, illustrating how the beams can partially shorten or stretch. Our findings show that increasing horizontal stiffness stiffens the system, leading to transitions from softening to linear to hardening behaviors near the second resonance frequency, indicating a bifurcation point. This highlights nonlinear Hencky's models' ability to handle complex beam behaviors that Euler-Bernoulli's theory cannot fully capture.

Looking ahead, future research could explore nonlinear studies of boundary conditions like varying slope angles, utilizing nonlinear Hencky's models to enhance our understanding of beam dynamics in diverse applications. Overall, despite their computational demands, Hencky's models offer substantial benefits in accurately analyzing complex beam behaviors.


\section*{Author contribution statement}
Mohammad Parsa Rezaei: Conceptualization, Investigation, Methodology, Validation, Software, Formal Analysis, Visualization, Writing – Original draft preparation, Writing – Reviewing and Editing; Grzegorz Kudra: Conceptualization, Investigation, Methodology, Validation, Software, Formal Analysis, Writing – Reviewing and Editing, Supervision; Mojtaba Ghodsi: Supervision, Investigation, Software, Formal Analysis; Jan Awrejcewicz: Supervision, Investigation, Formal Analysis.

\section*{Acknowledgements}
This study has been supported by the National Science Center, Poland under the grant PRELUDIUM 22 No. 2023/49/N/ST8/00823. This article was completed while the first author, Mohammad Parsa Rezaei, was a doctoral candidate at the Interdisciplinary Doctoral School at Lodz University of Technology, Poland. For the purpose of Open Access, the authors have applied a CC-BY public copyright license to any Author Accepted Manuscript (AAM) version arising from this submission.
This is an AAM of an article published by Elsevier in Journal of Sound and Vibration on 5 November 2024, available at: https://doi.org/10.1016/j.jsv.2024.118807.

\section*{Declaration of competing interest}
The authors certify that no personal relationships or known competing financial interests could have appeared to influence the research presented in this paper.
\section*{Data availability}
Data will be provided upon request.




\appendix 
\section{Building Matrices} \label{appendix:A}
This appendix provides a detailed representation of the matrices referenced in subsection~\ref{Matrix Formulation}. The elements of these matrices are reformulated according to the row index ($1 \leq i \leq n$), column index ($1 \leq j \leq n$), and the size of the matrix ($n$) for ease of coding implementation, as follows:
\begin{itemize}\label{eq:HenckeyMassMatrixUpperTriangleElements}
\item For $ i \leq j$ \&  $i =j ... n-1$:
\begin{align}
&M_{i,j} = \frac{(n - j) c_{i,j}}{n^3},  K_{i,i+1} = -n,  K_{i,i+j+1} = 0,&\nonumber \\ & N_{i,j} = \frac{(n-j) s_{i,j}}{n^3}, 
{K_x^{Total}}_{i,j} = 0,  {K_y}_{i,j} = 0, \nonumber \\& K_{\phi_{i,j}} = 0.&
\end{align}
\item For $ j \leq i$, \& $i =j ... n$:
\begin{align}
M_{j,i} &= M_{i,j},  & K_{j,i} &= K_{i,j},  & N_{j,i} &= -N_{i,j},  & K_{j,i} &= K_{i,j}, \nonumber \\
{K_x^{Total}}_{j,i} &= {K_x^{Total}}_{i,j}, & {K_y}_{i,j} &= {K_y}_{j,i}, & K_{\phi_{j,i}} &= K_{\phi_{i,j}}.
\end{align}
\item For $ i = j$ \& $i=1 ... n - 1$:
  \begin{align}
M_{i,i} &= \frac{6n-6i-1}{6n^3}, & K_{i,i} &= 2n, & N_{i,i} &= 0,\nonumber \\
{K_x^{Total}}_{i,i} &= s_i \left(\frac{2n-1}{2n^2} - {sc}_n\right), & {K_y}_{i,i} &= c_i {ss}_n, & K_{\phi_{i,i}} &= 0, \nonumber \\
 F_{i,1} &= \frac{(n-i)c_i}{n^2}.
\end{align}
\item For $ i = j=n$:
   \begin{align}
M_{n,n} &= \frac{1}{24n^3}, & K_{n,n} &= n, & N_{n,n} &= 0, \nonumber\\
{K_x^{Total}}_{n,n} &= \frac{1}{2}s_n \left(\frac{2n-1}{2n^2} - {sc}_n\right), & {K_y}_{n,n} &= \frac{1}{2}c_n {ss}_n, & K_{\phi_{n,n}} &= K_{\phi}, \nonumber \\
F_{n,1} &= \frac{c_n}{8n^2}.
\end{align}
\end{itemize}

To support our study, we have included supplementary materials with this paper. These materials contain MATLAB codes demonstrating the mathematical models and computations we used, specifically for this nonlinear Hencky’s beam model (NH). This code generates the last figure in our article (Fig.~\ref{figure16}). Changing the input parameters allows one to recreate other plots from our article using NH. We hope these resources help readers better understand NH and facilitate obtaining beam configuration results in future studies.

\end{document}